\documentclass[prb,onecolumn,aps,tightenlines,amssymb,amsmath,superscriptaddress]{revtex4}
\usepackage{graphicx}
\usepackage{bm}
\begin{document}
\title{Effect of electron-phonon interaction range on lattice polaron dynamics: a continuous-time quantum Monte Carlo study}
\author{P.~E.~Spencer}
\author{J.~H.~Samson}
\email{j.h.samson@lboro.ac.uk}
\affiliation{Department of Physics, Loughborough University, Loughborough LE11 3TU, United Kingdom}
\author{P.~E.~Kornilovitch}
\email{pavel.kornilovich@hp.com}
\affiliation{Hewlett-Packard Company, Mail Stop 321A, 1000 NE Circle blvd, Corvallis, Oregon 97330,
USA}
\author{A.~S.~Alexandrov}
\email{a.s.alexandrov@lboro.ac.uk}
\affiliation{Department of Physics, Loughborough University, Loughborough LE11 3TU, United Kingdom}
\begin{abstract}
We present the numerically exact ground state energy, effective mass, and isotope exponents of a
one-dimensional lattice polaron, valid for any range of
electron-phonon interaction, applying a continuous-time Quantum Monte Carlo (QMC) technique in a wide
range of coupling strength and adiabatic ratio. The QMC method is free from any systematic finite-size and
finite-time-step errors. We compare our numerically exact results with analytical
weak-coupling theory and with the strong-coupling $1/\lambda$ expansion. We show that the exact
results agree well with the canonical Fr\"ohlich and Holstein-Lang-Firsov theories in the weak and strong
coupling limits, respectively, for any range of interaction. We find a strong dependence of the polaron
dynamics on the range of interaction.  An increased range of interaction has a similar effect to an increased (less adiabatic) phonon frequency: specifically, a reduction in the effective mass.
\end{abstract}

\date{\today}

\maketitle


\section{Introduction}                                
While qualitative features of polarons were well recognized a long time ago and have been described in
several review papers and textbooks (see Ref.~\onlinecite{bry,mah,alemot,dev} for recent publications), there is
renewed interest in quantitative studies owing to the overwhelming evidence for polaronic carriers in
cuprates, fullerenes, and manganites (see, for example Ref.~\onlinecite{alemot,dev,mih,sal,alekabF,mil,alebra,zhao}).

Under certain conditions \cite{ale2} the multi-polaron system can be 
metallic but with polaronic carriers rather than bare electrons. There is 
a qualitative difference between the ordinary metal and the polaronic one. 
One can account for the electron-phonon (e-ph) interaction
in simple metals by applying Migdal's theorem \cite{mig}. The
theorem shows that the contribution of diagrams with ``crossing''
phonon lines (so called ``vertex'' corrections) is small if the
parameter $\lambda \hbar \omega/ E_{F}$ is small, where $\lambda$
is the dimensionless (BCS) e-ph coupling constant, $\omega$ is the
characteristic phonon frequency, and $E_{F}$ is the Fermi energy.
Neglecting the vertex corrections, Migdal calculated the
renormalized electron mass as $m^*=m_0(1+\lambda)$ (near the Fermi
level) \cite{mig}, where $m_0$ is the band mass in the absence of
electron-phonon interaction, and Eliashberg\cite{eli} extended
Migdal's theory to describe the BCS superconducting state at
intermediate values of $\lambda$, $\lambda \leq 1$. Later on many
authors applied Migdal-Eliashberg theory with $\lambda$ much
larger than 1 (see, for example, Ref.~\onlinecite{sca}).

On the other hand, starting from the infinite coupling limit,
$\lambda=\infty$ and applying the inverse ($1/\lambda$) expansion
technique \cite{fir} one can show \cite{ale,alemaz,ale0} that the
many-electron system collapses into the small polaron regime at
$\lambda \sim 1$ almost independently of the adiabatic ratio
$\hbar\omega/E_{F}$. This regime is beyond Migdal-Eliashberg
theory, where the effective mass approximation is used and the electron bandwidth is infinite. It is a well established theorem that a self-trapping crossover is
analytical in the coupling strength, so that one could believe
that the sum of all diagrams (including the vertex corrections)
should produce the exact solution if the expansion converges. Indeed, results of QMC
simulations based on summing the Feynman diagrams \cite{pro}
provide the exact answer in the continuous (large polaron) model.
On the other hand, the small polaron regime cannot be reached by
summation of the standard Feynman-Dyson perturbation diagrams
using a translation-invariant Green function
$G({\bf r},{\bf r'},\tau)=G({\bf r}-{\bf r'},\tau)$ with the Fourier
transform $G({\bf k}, \Omega)$ prior to solving the Dyson equations
on a discrete lattice.  This assumption excludes the possibility of
local  violation of the translational symmetry \cite{land} due to
the lattice deformation in any order of the Feynman-Dyson perturbation
theory similar to the absence of the anomalous (Bogoliubov) averages
in any order of perturbation theory \cite{mig}. One way to describe the formation of  the lowest polaronic band is to introduce an
infinitesimal translation-noninvariant potential, which should be set
zero only in the final solution obtained by the summation of Feynman
diagrams for the Fourier transform $G({\bf k},{\bf k'},\Omega)$ of
$G({\bf r},{\bf r'},\tau)$ rather than for $G({\bf k}, \Omega)$ \cite{alemaz}.
As in the case of the off-diagonal superconducting order parameter, the
off-diagonal terms of the Green function, in particular the Umklapp terms with ${\bf k'}={\bf k+G}$,
drive the system into a
small polaron ground state at sufficiently large coupling.  Setting the
translation-noninvariant potential  to zero in the solution of the
equations of motion restores the translation symmetry but in a polaron
band rather than in the bare electron band, which turns out to be an
excited state \cite{aleeur}. Alternatively, one can work with momentum eigenstates throughout
the whole coupling region, but taking into account the finite electron bandwidth (i.~e., including Umklapp terms).
  In recent years many such numerical and analytical
studies have confirmed the  conclusion \cite{ale} that the Migdal-Eliashberg theory breaks down at $\lambda \gtrsim 1$ (see
Refs.~\onlinecite{kab,kab2,bis,feh,mar,tak,feh2,tak2,rom,lam,zey,wag,aub,trugman,Korn2}
and references therein).

In ordinary metals, where the Migdal approximation is believed to be valid, 
the renormalized effective mass of electrons 
is independent of the ion mass $M$ because the electron-phonon interaction 
constant $\lambda$ does not depend on $M$. However, when the e-ph interaction 
is sufficiently strong, the electrons form polarons dressed by lattice 
distortions, with an effective mass $m^{\ast} = m_0 \exp (\gamma E_p/\hbar\omega)$. 
Here $E_p$ is the polaron binding energy (or the polaron shift), and $\gamma$ is a numerical constant 
that depends on the radius of electron-phonon interaction and is typically
less than 1. While $E_p$ in the above expression does not depend on the ion
mass, the phonon frequency does.  As a result, there is a large isotope effect 
on the carrier mass in polaronic conductors, $\alpha_{m} = (1/2)\ln (m^*/m)$ 
\cite{ale3}, in contrast to the zero isotope effect in ordinary metals. Such an 
effect was found experimentally in the cuprates \cite{zhao} and manganites 
\cite{pet}. A recent high resolution angle resolved photoemission spectroscopy 
study \cite{lan0} provided further compelling evidence for strong e-ph 
interaction in the cuprates. It revealed a fine phonon structure in the electron
self-energy of underdoped La$_{2-x}$Sr$_x$CuO$_4$ samples \cite{lan0,shencon}
and a complicated isotope effect in the electron spectral function of Bi2212 
that depended on the electron energy and momentum \cite{lan}.

With increasing phonon frequency
the range of validity of the $1/\lambda$ polaron expansion extends
to smaller values of $\lambda$ \cite{ale2}.  As a result, the
region of applicability of the Migdal-Eliashberg approach (even
with vertex corrections) shrinks to smaller values of the coupling,
$\lambda < 1$, with increasing $\omega$.  Strong correlations between
carriers might reduce this region further \cite{feh}. Carriers in the
fascinating novel materials are strongly coupled with high-frequency
optical phonons, making small polarons and  non-adiabatic effects
relevant for high-temperature superconductivity and colossal
magnetoresistance phenomena. Indeed the characteristic phonon energies
$0.05$--$0.2$ eV in cuprates, manganites and in doped fullerenes are
of the same order as the generally accepted values of the hopping
integrals $t \simeq 0.1$--$0.3$ eV \cite{Alex0}.

The continued interest in polarons extends beyond physical description of 
low-mobility conductors such as the oxides or doped polymers. The field has been a 
testing ground for analytical and numerical techniques for several decades.  In the 
past 25 years, several families of powerful numerical methods have been developed and 
successfully applied to one-, two-, and multi-polaron lattice models.  These are
the quantum Monte Carlo (QMC) simulations 
\cite{Blan,Hir,HirFra,Raedt,AlexRos,Kornil,pro,Hoh,Mac}, 
exact diagonalization of finite clusters \cite{kab2,mar,feh,stephan}, 
advanced variational methods \cite{dev,lam,rom,aub,trugman}, 
and the density-matrix renormalization group \cite{white}. 
Many methods have been developed so far as to enable reliable calculation of not only 
static and thermodynamic polaron properties, but also of the effective mass, spectrum,
and, in some cases, the spectral function of the polaron. 

At the same time, the bulk of the lattice polaron studies have been limited to the short-range electron-phonon interactions described by the Holstein model \cite{Hol}.
In numerical calculations, the locality of the interaction usually simplifies the 
algorithm and reduces the finite-size errors.  However, as pointed out by two
of us (AK)\cite{Korn2}, the Holstein model is {\em not} a typical but an 
{\em extreme} polaron model because the screening length is normally larger than the lattice constant in doped insulators.  
It yields {\em the highest possible} value of the polaron mass in the strong coupling limit, if  lattice vibrations are isotropic or polarised perpendicular to the hopping direction \cite{ku}.
   With an on-site 
electron-phonon interaction, during every polaron hop the existing lattice deformation
has to relax completely to the undeformed state, while a full deformation has to form 
again at the new location of the particle.  Such a process results in the exponentially
small overlap between the initial and the final states, and in an exponentially large
effective mass with $\gamma = 1$.  Real ionic solids with low density of free carriers
are characterized by poor screening and are more appropriately described by a long-range
electron-phonon interaction.  Thus the {\em lattice Fr\"ohlich model} introduced in 
Ref.~\onlinecite{Korn2} is intermediate between the extremes of the Holstein and Fr\"ohlich
 \cite{fro} limits.  On one hand, it is a lattice model (like the Holstein one), and the ratio of the
hopping integral to the phonon frequency is an important parameter.  On the other hand,
the electron-phonon interaction is long-range, as in the Fr\"ohlich model.  It was 
shown in Ref.~\onlinecite{Korn2} that in this intermediate case the polaron mass      
still grows exponentially with the polaron binding energy $E_p$ but the parameter 
$\gamma$ is now less than unity.  That leads to much reduced numerical values of the
polaron mass, hence the term {\em mobile small polaron}.  The model was further studied
by numerical cluster diagonalization \cite{feh3} and $1/\lambda$ expansion 
\cite{ale2,alechan}.  In addition, the {\em two}-particle model with non-local 
electron-phonon interactions was studied variationally \cite{bon2} and by   
the $1/\lambda$ expansion technique \cite{ale2,alekorJ}.  These studies confirmed the
original conclusion that a long-range interaction significantly reduces the effective 
mass of the carrier, polaron or bipolaron, sometimes by several orders of magnitude,
in comparison with the Holstein model.  It also makes the self-trapping transition
more gradual as a function of $\lambda$, and better describable by the Lang-Firsov
theory \cite{fir}.  These findings are in agreement with some earlier studies on
long-range interactions in narrow-band models \cite{eag}.   
 
In this paper, we further generalize the lattice Fr\"ohlich model of Ref.~\onlinecite{Korn2}
to electron-phonon interaction of some finite radius $R$.  We perform a systematic
study of the {\em single} polaron problem in one dimension as a function of $R$.
In the local limit $R \rightarrow 0$ we recover the results of the Holstein model
obtained in the past by various methods mentioned above.  In the infinite-$R$ limit
the original AK model and its results are fully recovered as well.  Our computational
tool will be the continuous-time path-integral quantum Monte Carlo algorithm developed
previously by one of us \cite{Kornil}.  This method is particularly suited for 
investigating long-range electron-phonon interactions because the phonon degrees of
freedom are integrated out analytically.  Thus the shape of the interaction does not
complicate the algorithm at all, but simply modifies the weight function of a Monte
Carlo configuration.  The method works on infinite lattices and in arbitrary dimensions,
eliminating finite-size errors; there is also no truncation of the phonon Fock space.  The method is also free from finite-time-step errors
because it is formulated in continuous time.  The method enables unbiased calculation
(i.e. no numerical errors besides statistical fluctuations) of the polaron energies,
effective mass, spectrum, density of states, isotope exponents, the number of excited 
phonons, and other quantities.  

In addition to presenting novel results on the finite-radius Fr\"ohlich model, we use
the present paper to explain many technical details of the polaron QMC method \cite{Kornil},
which have not previously been published.  The electron-phonon Hamiltonian is introduced 
in section II.  In section III we describe the continuous-time Monte Carlo method. 
In sections IV--VI we present the numerical results for the energy, effective mass, 
number of dressing phonons, and isotope exponents of lattice polarons for different
$R$, and compare them with weak-coupling and strong-coupling analytical results and with numerical results of other authors.  Section VII summarizes our conclusions.

\section{Electron-phonon model}         \label{sect2}  
\subsection{General model Hamiltonian}

The electron-phonon model under investigation represents a single electron interacting with {\it all} the
ions of an infinite hypercubic lattice, with one vibrational degree of freedom per unit cell. The
Hamiltonian takes the form 
\begin{equation}
    H = H_{\rm e}+H_{\rm ph}+H_{\rm e-ph}
    \label{MHh},
\end{equation}
where
\begin{equation}
    H_{\rm e}=-t\sum_{\langle{\bf nn'}\rangle}c_{\bf n}^{\dag}c_{\bf n'}
    \label{MHhe},
\end{equation}
\begin{equation}
    H_{\rm ph}= \frac{1}{2M}\sum_{\bf m}P_{\bf m}^{2}
        +\frac{M\omega^{2}}{2} \sum_{\bf m}\xi_{\bf m}^{2}
    \label{MHhph},
\end{equation}
and
\begin{equation}
        H_{\rm e-ph}= - \sum_{\bf nm}f_{\bf m}({\bf n})
    c_{\bf n}^{\dag}c_{\bf n}\xi_{\bf m}
    \label{MHheph}.
\end{equation}

The {\it free-electron term} $H_{\rm e}$ describes the movement of a single electron through the lattice by
the process of nearest-neighbor hopping. Here the operator $c_{\bf n}^{\dag}$ creates an electron on site
${\bf n}$, the operator $c_{\bf n'}$ destroys an electron on site ${\bf n}'$, and $\langle{\bf nn'}\rangle$
denotes pairs of nearest-neighbor sites.
The {\it phonon term} $H_{\rm ph}$ represents the vibrations of the lattice ions.
Here the operator $\xi_{\bf m}$ is the displacement of the ${\bf m}$th ion from its
equilibrium position, and $P_{\bf m}=-i\hbar\partial /\partial\xi_{\bf m}$ its
momentum. It is assumed that the ions, each of ionic mass $M$, are non-interacting and
so have the same characteristic (phonon) frequency $\omega$.
The final part of the Hamiltonian, the electron-phonon term $H_{\rm e-ph}$, is of the
``density-displacement'' type, where the interaction energy between the electron and
the ${\bf m}$th ion is proportional to  $\xi_{\bf m}$ (the displacement of the ${\bf m}$th
ion from its equilibrium position).  Here $c_{\bf n}^{\dag}c_{\bf n}$ is the electron
number operator, and $f_{\bf m}({\bf n})$ is interpreted as the {\it interaction force}
between the electron on site ${\bf n}$ and the ${\bf m}$th lattice ion.

The model is parameterized by two dimensionless quantities. The first is the {\it dimensionless phonon
frequency}
\begin{equation}
    \bar{\omega}=\hbar\omega/t
    \label{MHomega}.
\end{equation}
The second is related to the small-polaron binding energy $E_{\rm p}$, derived in section (\ref{sect6}),
which serves as a natural and convenient measure of the strength of the electron-phonon interaction. The
dimensionless {\it electron-phonon coupling constant} is defined as
\begin{equation}
    \lambda = \frac{E_{\rm p}}{zt}
    =\frac{1}{2M\omega^{2}zt}\sum_{\bf m}f_{\bf m}^{2}(0)
     \label{Plam},
\end{equation}
where $zt$ is the bare-electron half bandwidth, with $z$ the lattice coordination number.

\subsection{Discrete Fr\"{o}hlich model}
\begin{figure}
    \centering
\centerline {
\includegraphics[viewport=0 50 600 200,clip]{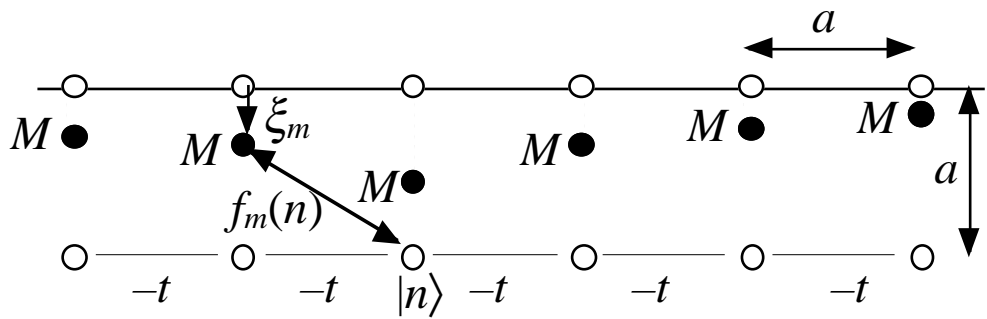}
}
    \caption{Geometry of the Fr\"{o}hlich model (\ref{MHh}--\ref{MHheph}) shown in one
    dimension.  The mobile charge carrier moves on the lower chain with nearest-neighbor hopping integral $t$ and
    interacts with all the ions of the upper chain.  The displacements $\xi_m$ of
    the ions are polarized in a direction perpendicular to the
    chains.}
    \label{polaron}
\end{figure}

Some time ago Alexandrov and Kornilovitch  proposed a long-range
discrete Fr\"{o}hlich interaction \cite{Korn2} to describe the
interaction between a hole and the apical oxygen ions in
high-$T_{\rm c}$ superconducting materials. The model is
depicted in FIG.~ \ref{polaron} for the one-dimensional case. The
mobile carrier (electron or hole) may hop from site to
nearest-neighbor site along the lower chain. The chain consists of
an infinite number of lattice sites with lattice constant $a$. The
electron interacts with {\it all} the ions which reside at the
lattice sites of a similar chain that is parallel to the first.
The separation of the two chains is equal to the lattice constant
$a$.  We assume that the vibrations of the ions are polarized in a
direction that is perpendicular to the chains, and that the ions
do not interact with each other.

Let us find the appropriate form for the interaction force $f_m(n)$ between the mobile
charge-carrier on the $n$th site (of the lower chain) and the $m$th ion (of the upper chain). Since
both $m$ and $n$ are measured in units of $a$, we choose from this point on to take $a=1$. The
presence of the charge-carrier displaces the $m$th ion by a {\it small} distance $\xi_m$ in a
direction perpendicular to the chain, as shown in FIG.~\ref{polaron}. By expanding the Coulomb potential in powers of $\xi_m$, we deduce\cite{Korn2} that the Hamiltonian for the
discrete Fr\"{o}hlich model is that of our generalized model Hamiltonian, Eq.~\ref{MHh}, with the
electron-phonon interaction force having the form
\begin{equation}
    f_{m}({n})=\frac{\kappa}{\left[ (m-n)^{2}+1 \right]^{3/2}}
    \label{FMf}
\end{equation}
with a constant $\kappa$. Physically, this model was proposed in
order to represent the interaction between a hole in the
copper-oxygen layer (lower chain) and the apical oxygens in the
ionic layer (upper chain) contained within the structure of
certain doped high-${T_{\mathrm c}}$ superconductors such as
$\mathrm{YBa_{2}Cu_{3}O}_{6+x}$ \cite{Korn2}. These materials are
highly anisotropic due to the fact that the holes are sharply
localized in the copper-oxygen layer, giving rise to poor
conduction in the $c$-direction (normal to copper-oxygen layer).
This leads to very poor screening of the electron-phonon
interaction in the $c$-direction, and almost complete screening in
the $a$-$b$ plane.  This justifies the restriction to phonon modes
polarized in the $c$ direction.

\subsection{Screened Fr\"{o}hlich model}
Our aim in this paper is to investigate the way in which the {\it shape} of the long-range electron-phonon
interaction affects the properties of the polaron. It is therefore interesting to study the {\it screened}
Fr\"{o}hlich model, in which the screening effect due to the presence of other electrons in the lattice is
taken into account from within $f_{m}({n})$.  Accordingly, let us define the interaction force for
the {\it screened Fr\"{o}hlich model} as
\begin{equation}
    f_{m}({n})=\frac{\kappa}{\left[ (m-n)^{2}+1 \right]^{3/2}}
    \exp\left(-\frac{|m-n|}{R_{\rm sc}}\right)
    \label{FMfsc},
\end{equation}
where $R_{\rm sc}$ is the screening length. That is, the screened force is the unscreened force multiplied
by an exponential damping factor. Increasing the value of $R_{\rm sc}$ decreases the screening effect and
thus increases the width of the interaction force.

The Holstein model describes an electron that interacts only with
the oscillator it currently occupies (``short-range''
interaction).  This may be regarded as a special case of equation
(\ref{FMfsc}) with $R_{\rm sc}\to 0$, so that
$f_{m}({n})=-\kappa\delta_{mn}$.  Simply by altering the value of the parameter $R_{\rm sc}$, we can easily cross over from the Holstein
model, through the screened Fr\"{o}hlich model, to the unscreened Fr\"{o}hlich model, in a universal
manner. In this paper we consider the following four cases in one dimension:

\begin{enumerate}
    \item Holstein model with $R_{\rm sc}\to 0$.
    \item Screened Fr\"{o}hlich model, with $R_{\rm sc}=1$.
    \item Screened Fr\"{o}hlich model, with $R_{\rm sc}=3$.
    \item Unscreened Fr\"{o}hlich model with $R_{\rm sc}\to \infty$.
\end{enumerate}
\begin{figure}
\centerline {
\includegraphics[scale=0.5, viewport=0 360 800 800,clip]{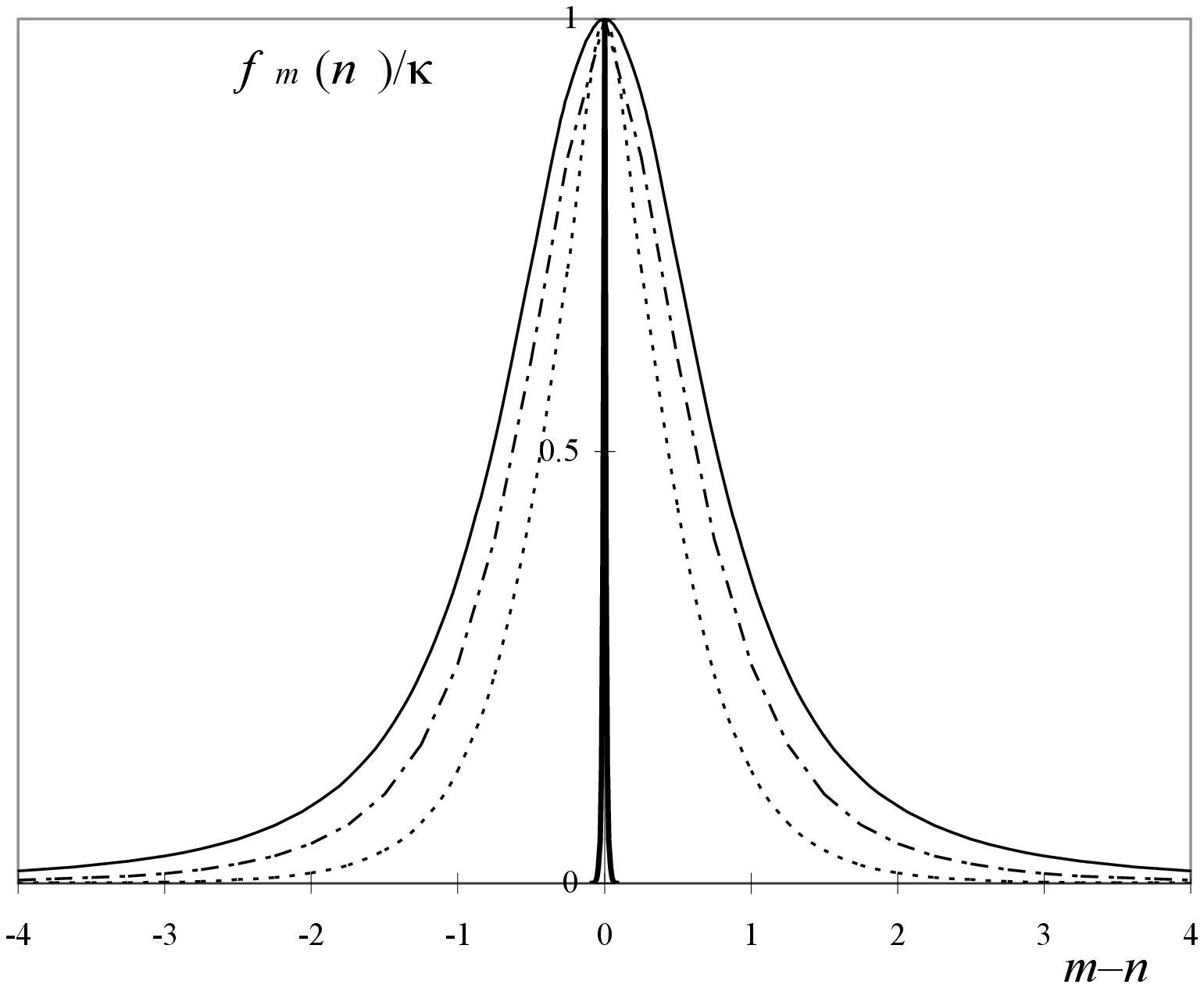}
}
\caption{The shape of the screened Fr\"{o}hlich interaction force,
Eq.~\ref{FMfsc}, at screening lengths of $R_{\rm sc}\to 0$
(bold line, Holstein interaction), $R_{\rm sc}=1$ (dashed line), $R_{\rm sc}=3$ (dot-dashed line) and $R_{\rm
sc}\to\infty$ (thin line, non-screened Fr\"{o}hlich interaction).}
\label{forces}
\end{figure}

The shapes of the electron-phonon interaction force $f_m(n)$ for
each of the above screening lengths are shown in FIG.~\ref{forces}.
Note that, based on calculations involving the dynamic properties of
the polaron response \cite{ale2}, the amount of screening we impose
here is greater than that expected in the high-$T_{\rm c}$ compounds.

\section{Path integral approach} \label{sect3}
\subsection{Effective mass using a partial partition function}
\label{secteff}
The effective mass of the polaron $m^{*}$ is defined for the
isotropic or one-dimensional case as \cite{Kornil}
\begin{equation}
    (m^*)^{-1}= \lim_{P\rightarrow 0}
    \frac{\partial^{2} E(P)}{\partial P^2},
    \label{PP:Mm1}
\end{equation}
where $E({\bf P})$ is the ground state energy for {\it total momentum} ${\bf P}$
(sum of the momenta of the electron and {\it all} the phonons). 

The evaluation of $(m^*)^{-1}$ by differentiating QMC energies is
not practical within our approach because a minus-sign problem arises for finite
momentum, exacerbating the errors already present in such a
procedure. The usual means of extracting dynamical properties
(such as the effective mass) from QMC simulation is by making use
of some kind of analytical continuation from imaginary to real
time. However, it is possible to infer $m^{*}$ {\it directly} from
QMC simulation by considering electron trajectories with {\it
twisted} (rather than periodic) boundary conditions in imaginary
time. Kornilovitch\cite{Kornil} showed (for the isotropic
or one-dimensional case) that
\begin{equation}
   \frac{m_0}{m^{*}} = \lim_{\beta\rightarrow\infty}\frac{1}{2\beta ta^{2}}\frac
    {\sum_{\Delta{\bf r}}(\Delta{\bf r})^2 Z_{\Delta{\bf r}}}
    {\sum_{\Delta{\bf r}}Z_{\Delta{\bf r}}},
    \label{EMmm1}
\end{equation}
where $m_0 = \hbar^2/(2ta^2)$ is the bare electron mass, and
\begin{equation}
    Z_{\Delta {\bf r}}=\int d^{N}\bm\xi
     \langle \{\xi_{{\bf m}+  {\bf r'}- {\bf r}}\},{\bf r'}
    \mid e^{-\beta H}\mid \{  \xi_{\bf m}\},{\bf r}\rangle
    \label{PP:Zr},
\end{equation}
is a ``partial partition function'' (which is similar in form to
the total partition function of the system), with
\begin{equation}
    \int d^{N}\bm\xi=
    \prod_{{\bf m}=1}^{N}\left[\int_{-\infty}^{\infty}d\xi_{{\bf m}}\right]
    \label{PP:Zr2}.
\end{equation}
Here $|{\bf r}\rangle$ is the electron basis, $|\{\xi_{\bf
m}\}\rangle = |\xi_{1},\xi_{2},\xi_{3},\cdots,\xi_{N}\rangle$ is
the ionic displacement basis, and the summations over $\Delta {\bf r}$
includes all possible values of $\Delta {\bf r}={\bf r}' -{\bf
r}$.

Given Eq.~\ref{EMmm1}, the effective mass may be obtained
from QMC simulation by taking the statistical average of
$(\Delta{\bf r})^2$, sampled over trajectories of the path
integral formed from $Z_{\Delta {\bf r}}$. The dissimilarity
between the ``bra'' and ``ket'' states in Eq.~\ref{PP:Zr} produces
a path integral of $Z_{\Delta {\bf r}}$ having twisted (rather
than the usual periodic) boundary conditions. Note that, since we
need only consider the case of ${\bf P}={\bf 0}$, there is no sign
problem.

\subsection{Continuous imaginary time}
QMC schemes have recently been developed that are implemented
directly in {\it continuous} imaginary time for lattice
models\cite{pro,Beard,Mac}, eliminating the problematic
finite-time-step error associated with the traditional
discrete-time approach. The partial partition function $Z_{\Delta
{\bf r}}$ in Eq.~\ref{PP:Zr} is given in continuous-imaginary-time
path-integral form as\cite{Kornil}
\begin{equation}
    Z_{\Delta {\bf r}}=\int_{\rm tw}\mathcal{D}\bm\xi\mathcal{D}{\bf r}
    \exp (S)
    \label{PP:Zr3},
\end{equation}
where the phonon action reads
\begin{equation}
    S = \sum_{\bf m} S_{\bf m} =
    \sum_{\bf m}\int_{0}^{\beta} d\tau \left[
    -\frac{M}{2\hbar^{2}}\dot\xi_{\bf m}^{2}(\tau)
    -\frac{M\omega^{2}}{2}\xi_{\bf m}^{2}(\tau)
    +f_{\bf m}({\bf r}(\tau))\xi_{\bf m}(\tau)    \right]
    \label{CTaction},
\end{equation}
with $\dot\xi_{\bf m}(\tau)=\partial\xi_{\bf
m}(\tau)/\partial\tau$. In forming the path integral above, an
imaginary-time dimension $\tau$ has been introduced, having the
range $0\le\tau\le\beta$. The electron and phonon coordinates are
represented as continuous functions of imaginary time, ${\bf
r}(\tau)$ and ${\bf \xi_{\bf m}}(\tau)$, which can be interpreted
as continuous trajectories in $\tau$.

The symbol $\int_{\rm tw}$ in Eq.~\ref{PP:Zr3} represents the
integration over all possible trajectories under twisted boundary
conditions in imaginary time. The ``end states'' of the individual
trajectories, which are identified with the states $\langle
\{\xi_{{\bf m}+ {\bf r'}- {\bf r}}\},{\bf r'} \mid$ and $\mid \{
\xi_{\bf m}\},{\bf r}\rangle$ in Eq.~\ref{PP:Zr}, are given
by\cite{Kornil}
\begin{eqnarray}
    \mid \{\xi_{\bf m}(0)\}, {\bf r}(0)\rangle
    &=&
    \mid \{\xi_{\bf m}\},{\bf r}\rangle
    \nonumber \\
    \mid \{\xi_{{\bf m}}(\beta)\}, {\bf r}(\beta)\rangle
    &=&
    \mid \{\xi_{{\bf m}+ \Delta{\bf r}}\},{\bf r}+ \Delta{\bf r}\rangle
    \label{TBCbc0},
\end{eqnarray}
that is, the final state ($\tau=\beta$) is the initial state
($\tau=0$) with all the coordinates (electron and {\it all}
phonons) shifted by $\Delta{\bf r}$.

We may decompose the trajectory $\xi_{\bf m}(\tau)$ in
Eq.~\ref{CTaction} into the sum of a the classical path (the
trajectory that extremizes $S_{\bf m}$) and a deviation (or
``quantum fluctuation'') from it.  The part of $S_{\bf m}$ that
contains no terms involving quantum fluctuation is the classical
action. The classical action is an important quantity, and is
given by \cite{Feyn2,Kornil}
 \begin{eqnarray}
     S^{\rm cl}_{\bf m} =
    \frac{M\omega}{2\hbar\sinh(\hbar\omega\beta)}\left\{
 -[\xi_{\bf m}^{2}(0)+{\xi}_{\bf m}^{2}(\beta)]\cosh(\hbar\omega\beta)
    +2\xi_{\bf m}(0)\xi_{\bf m}(\beta) \right\}
+   \xi_{\bf m}(0)B_{\bf m}(\tau) + \xi_{\bf m}(\beta) C_{\bf
m}(\tau) 
     \nonumber  \\
 \hspace{5.0cm}{}+\frac{\hbar^{2}}{2M}\int_{0}^{\beta} \!\!\!
 \int_{0}^{\beta} \!\!\! d\tau d\tau'
 f_{\bf m}({\bf r}(\tau)) G(\tau,\tau') f_{\bf m}({\bf r}(\tau')),
 \label{PBCaepts}
\end{eqnarray}
where
\begin{equation}
    B_{\bf m}(\tau) \equiv \int_{0}^{\beta} \!\!\! d\tau
    \frac{\sinh(\hbar\omega(\beta-\tau))}
         {\sinh(\hbar\omega\beta)}f_{\bf m}({\bf r}(\tau))
    \label{PBCb},
\end{equation}
\begin{equation}
    C_{\bf m}(\tau) \equiv \int_{0}^{\beta} \!\!\! d\tau
    \frac{\sinh(\hbar\omega\tau) }
         {\sinh(\hbar\omega\beta)}f_{\bf m}({\bf r}(\tau))
    \label{PBCc},
\end{equation}
and the Green function is
 \begin{equation}
   G(\tau,\tau')=\frac{1}{\hbar\omega \sinh(\hbar\omega\beta)}
    \left\{ \begin{array}{ll}
     \sinh (\hbar\omega\tau) \sinh [\hbar\omega(\beta - \tau')],&
     \mbox{$0<\tau<\tau'$} \\
     \sinh [\hbar\omega (\beta - \tau)] \sinh (\hbar\omega\tau'),&
\mbox{$\tau'<\tau<\beta$}
                \end{array} \right.
    \label{CT:gf}.
 \end{equation}
Note that the phonon coordinates in Eq.~\ref{PBCaepts} are those
of the end-points only: $\xi_{\bf m}(0)$ and $\xi_{\bf m}(\beta)$.

\subsection{Analytical Phonon Integration}
We wish to integrate out the phonon degrees of
freedom from the problem analytically, that is, perform the {\it phonon path
integral}
\begin{equation}
   I_{\rm tw} = \int_{\rm tw}\mathcal{D}\bm\xi
    \exp\left(\sum_{\bf m}S_{\bf m}\right)
    =
    c_{\rm tw}\int_{\rm tw}d\bm\xi
    \exp\left(\sum_{\bf m}S^{\rm cl}_{\bf m}\right)
    \label{PBCi},
\end{equation}
where the non-classical part of $S$ (terms involving quantum
fluctuation) integrates to an unimportant constant\cite{Kornil}
$c_{\rm tw}$, reducing the problem to the integration of the
classical action $S^{\rm cl}_{\bf m}$. The integration must be
performed under twisted boundary conditions in imaginary time.
Accordingly, we impose the constraints
\begin{equation}
    \xi_{\bf m}(0) = \xi_{{\bf m}}
    \quad ;\quad
    \xi_{\bf m}(\beta) = \xi_{{\bf m}-\Delta{\bf r}}
        \label{Gtwbc}
\end{equation}
on $S^{\rm cl}_{\bf m}$ in Eq.~\ref{PBCaepts}, which one can see
produces mixed variable terms involving $\xi_{\bf m}\xi_{{\bf
m}-\Delta{\bf r}}$. The phonon integration cannot directly be
performed in this form.  However, we may proceed by transforming
$\xi_{\bf m}$ into real Fourier
components $a_{\bf q}$ and $b_{\bf q}$:
\begin{equation}
    \xi_{\bf m}=\frac{1}{\sqrt{DN}}\!\sum_{\bf q}
    (a_{\bf q}+ib_{\bf q})e^{i{\bf q\cdot m}}
    \label{Gxi},
\end{equation}
where $D$ is the dimensionality of the lattice and $DN$ is the
total number of phonon degrees of freedom. In this representation,
the transformed action $S^{\rm cl}_{\bf q}$ is diagonal, and so the phonon path integral in Eq.~\ref{PBCi}
decomposes to the product of single variable integrals according
to
\begin{equation}
    I_{\rm tw} = c_{\rm tw}\prod_{\bf q}\int_{\rm tw}da_{\bf q}db_{\bf q}
    \exp \left( S^{\rm cl}_{\bf q} \right)
    \label{GUab}.
\end{equation}
After performing the Gaussian integration (in $a_{\bf q}$ and
$b_{\bf q}$) the result is
\begin{equation}
    I_{\rm tw} = c_{\rm tw}\left[ \frac{\pi\hbar\sinh(\hbar\omega\beta)}
    {M\omega [ \cosh(\hbar\omega\beta)-\cos({\bf q}\cdot\Delta{\bf r}) ] }
     \right]^{DN/2} \exp A
    \label{GUab2},
\end{equation}
where
\begin{eqnarray}
    \lefteqn{A =
    \frac{\hbar\sinh(\hbar\omega\beta)\sum_{\bf m}
    B_{\bf m}(C_{{\bf m}-\Delta{\bf r}}-C_{\bf m})}
    {2M\omega[\cosh(\hbar\omega\beta)-\cos({\bf q}\cdot\Delta{\bf r})]}
    +\frac{\hbar\sinh(\hbar\omega\beta)\sum_{\bf m}
     (B_{\bf m}+C_{{\bf m}})^{2}}
    {4M\omega[\cosh(\hbar\omega\beta)-\cos({\bf q}\cdot\Delta{\bf r})]}
    }\hspace{0cm}\nonumber \\
    & &\hspace{4.6cm} +\sum_{\bf
m}\frac{\hbar^{2}}{2M}\int_{0}^{\beta}\!\!
   \int_{0}^{\beta}\!\! d\tau d\tau'
 f_{\bf m}({\bf r}(\tau)) G(\tau,\tau') f({\bf r}(\tau')),
    \label{Ga5}
\end{eqnarray}
which does not contain any phonon degrees of freedom.  We have
thus transformed the problem from that of an electron interacting
with many phonons to that of an electron with retarded
self-interaction, which allows the QMC method to be applied
effectively.

\subsection{Low temperature limit}
The result above may be conveniently rendered into the required
low temperature limit using $\cosh\hbar\omega\beta \approx
\sinh\hbar\omega\beta \approx \frac{1}{2}e^{\hbar\omega\beta}\gg
1$, to give
\begin{equation}
    I_{\rm tw}
    \propto \exp A=\exp\left(A_{\rm per}+\Delta A \right)
    \label{GUlim},
\end{equation}
where
\begin{equation}
    A_{\rm  per}=\frac{z\lambda\bar{\omega}}{2\Phi_{0}(0,0)}
     \int_{0}^{\bar{\beta}} \!\!\!
     \int_{0}^{\bar{\beta}} \!\!\!
     d\bar{\tau} d\bar{\tau}'
     e^{-\bar{\omega}\frac{\bar{\beta}}{2}}\left(
     e^{\bar{\omega}(\frac{\bar{\beta}}{2}-|\bar{\tau}-\bar{\tau}'|)}
     + e^{-\bar{\omega}(\frac{\bar{\beta}}{2}-|\bar{\tau}-\bar{\tau}'|)}
     \right)
     \Phi_{0}\left({\bf r}(\bar{\tau}),{\bf r}(\bar{\tau}')\right)
    \label{CTaper}
\end{equation}
is the low temperature action for periodic boundary conditions,
and
\begin{equation}
    \Delta A=\frac{z\lambda\bar{\omega}}{\Phi_{0}(0,0)}
     \int_{0}^{\bar{\beta}} \!\!\!
     \int_{0}^{\bar{\beta}} \!\!\!
     d\bar{\tau} d\bar{\tau}'
     e^{-\bar{\omega}\bar{\tau}}e^{-\bar{\omega}(\bar{\beta}-\bar{\tau}')}
     \left[
     \Phi_{\Delta{\bf r}}\left({\bf r}(\bar{\tau}),{\bf r}(\bar{\tau}')\right)
     -\Phi_{0}\left({\bf r}(\bar{\tau}),{\bf r}(\bar{\tau}')\right)
     \right]
    \label{CTdela}
\end{equation}
is the correction for twisted boundary conditions, in
dimensionless form.  Here $\bar{\tau}=t\tau$ and
$\bar{\beta}=t\beta$ define dimensionless imaginary time; the
parameters $\bar{\omega}$ and $\lambda$ are defined in
Eqs.~\ref{MHomega} and \ref{Plam} respectively; and the {\it
lattice summation} is defined as
\begin{equation}
    \Phi_{\Delta{\bf r}}\left({\bf r}(\bar{\tau}),{\bf r}(\bar{\tau}')
    \right)=\sum_{\bf m}\bar{f}_{\bf m}({\bf r}(\bar{\tau}))
    \bar{f}_{{\bf m}+\Delta{\bf r}}({\bf r}(\bar{\tau}'))
    \label{CTphi2}.
\end{equation}
Note that the dimensionless quantity $\bar{f}_{\bf m}({\bf n})$
represents the {\it shape} or {\it form} of the electron-phonon
interaction force, defined via the decomposition
\begin{equation}
    f_{\bf m}({\bf n})=\kappa\bar{f}_{\bf m}({\bf n})
    \label{CTft},
\end{equation}
where $\kappa=\left[2z\lambda M
t^{3}\bar{\omega}^{2}/\left(\hbar^{2}\sum_{\bf m}\bar{f}^{2}_{\bf
m}(0)\right)\right]^{1/2}$ takes the dimensions of force.

\section{Continuous-time Monte Carlo}   \label{sect4}   
\subsection{Algorithm}  \label{algo}
Traditionally, path-integral QMC simulation is implemented in {\it
discrete} imaginary-time, where the trajectory is represented by
the position of the electron in each of a large number of
imaginary-time slices.  The use of discrete time introduces the
problematic finite-time-step {\it systematic} error, which scales
with the square of the time-slice-width.

\begin{figure}
\includegraphics[scale=0.7, viewport=0 480 800 800,clip]{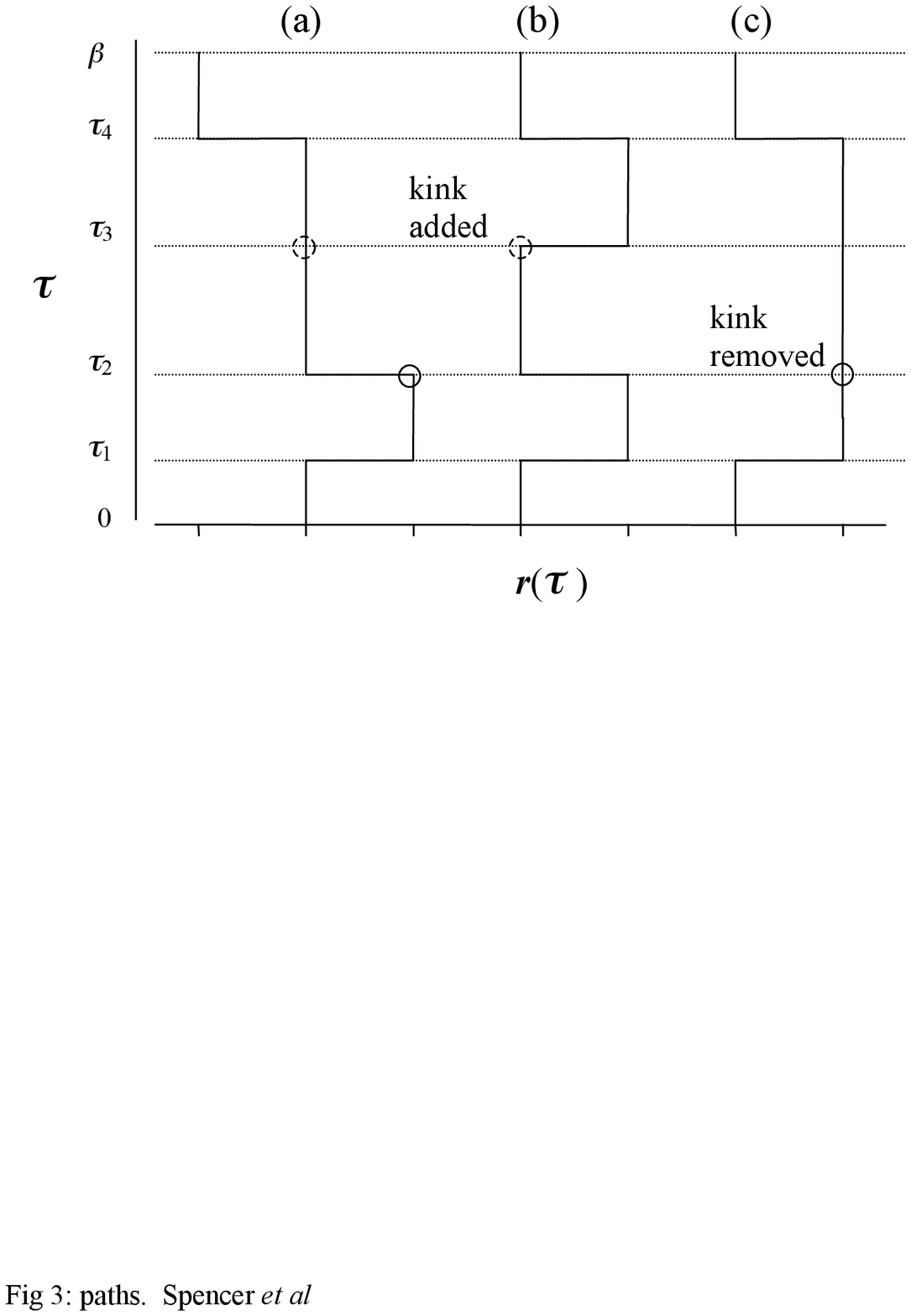}
    \caption{Illustration of a one-dimensional electron trajectory in imaginary
time. The point in imaginary time at which the electron hops to a
neighboring site is known as a kink.  (a) Shows a trajectory with three
kinks: occurring at $\tau_{1}$, $\tau_{2}$, and $\tau_{4}$.  (b) The same
trajectory, but with a kink added at time $\tau_{3}$. The entire trajectory
above $\tau_{3}$ is shifted by one lattice parameter.  (c) The same trajectory
as in (a) but with the kink at $\tau_{2}$ removed. Again, the entire
trajectory above $\tau_{2}$ is shifted accordingly.}
    \label{paths}
\end{figure}

A path-integral QMC scheme implemented directly
in {\it continuous} imaginary time has been developed for systems
with a discrete  basis \cite{pro, Beard}. Here, the electron
trajectory is represented as finite intervals of imaginary time in
which the system remains in a particular state, separated by
sporadic transitions from one state to another (an electron hop).
The points in imaginary time at which the state of the system
changes are called ``kinks'', as shown in FIG.~ \ref{paths}. It is necessary to consider the
statistics governing different {\it directions} of kink {\it
independently} of one another. For our one-dimensional case with
nearest-neighbor hopping, we need only consider single {\it left}
and {\it right} kinks. The use of continuous time completely
eliminates the finite-time-step error, rendering the scheme
``numerically exact".

If $N_{s}$ is the number of kinks of direction $s$, we wish to
generate random states according to the Monte Carlo weight
\begin{equation}
    w(\{N_{s}\})=w_{\rm el}(\{N_{s}\}) \ \ w_{\rm ph}(\{N_{s}\})
    \label{CQwsp},
\end{equation}
where the weight from the electron subsystem
\begin{equation}
   w_{\rm el}(\{N_{s}\}) = \prod_s\frac{(t\beta)^{N_{\!s}}e^{-t\beta}}{N_{\!s}!}
    \label{CTAp2}
\end{equation}
is given by the Poisson distribution, and the phonon-induced
weight
\begin{equation}
    w_{\rm ph}(N_{s})=\exp[A(N_{\!s})]
    \label{CTAp}
\end{equation}
is given by Eq.~\ref{GUlim}.

Proposed changes to the shape of the trajectory are generated by
the addition or deletion of single kinks. This is sufficient in
practice. In order to increase efficiency one might also consider
changing a kink-direction, repositioning a kink in imaginary-time,
or altering the temporal ordering of the kinks. The Metropolis
method \cite{Met,Bind} accepts or rejects the trial change from
state $\mu$ to state $\mu'$ with a transition probability
$P(\mu\rightarrow\mu')=g(\mu\rightarrow\mu')a(\mu\rightarrow\mu')$,
where $g(\mu\rightarrow\mu')$ is the {\it sampling distribution}
and $a(\mu\rightarrow\mu')$ is the {\it acceptance probability}.
For the case of $N_{\!s}\geq 1$ (one or more kinks exist), we
choose $g(N_{\!s}+1\rightarrow N_{\!s})=g(N_{\!s}\rightarrow
N_{\!s}+1)=1/2$, and so the acceptance probability is given by
\begin{eqnarray}
    a_{\rm add}(N_{\!s}\rightarrow N_{\!s}+1)  &=&
    \min\left\{ 1,
    \frac{g(N_{\!s}+1\rightarrow N_{\!s})}{g(N_{\!s}\rightarrow N_{\!s}+1)}\frac{W(N_{\!s}+1)}{W(N_{\!s})}
    \right\}
    \label{MCm} \\
     &=& \min \left\{ 1,\frac{t\beta}{N_{\!s}+1} \exp\left[A(N_{\!s}+1)-A(N_{\!s})\right]\right\}
    \label{CTA1ad}
\end{eqnarray}
to add a kink, and similarly
\begin{equation}
    a_{\rm rem}(N_{\!s}+1\rightarrow N_{\!s})=
    \min \left\{ 1,\frac{N_{\!s}+1}{t\beta} \exp[A(N_{\!s})-A(N_{\!s}+1)]\right\}
    \label{CTA1re}
\end{equation}
to remove a kink of direction $s$.  For the case of $N_{\!s}=0$, we can
only add a kink, and so $g(0\rightarrow 1)=1$, which gives
\begin{equation}
    a_{\rm add}(0\rightarrow 1)=
    \min \left\{ 1,\frac{t\beta}{2} \exp[A(1)-A(0)]\right\}
  \label{CTA2ad}
\end{equation}
and
\begin{equation}
    a_{\rm  rem}(1\rightarrow 0)=
    \min \left\{ 1,\frac{2}{t\beta} \exp[A(0)-A(1)]\right\}
    \label{CTA2re}.
\end{equation}

The continuous-imaginary-time QMC step used in this work has the
following structure:
\begin{enumerate}
    \item  Randomly select a kink direction $s$ for the trial change. In the case of a
    one-dimensional system, this is left or right.

    \item  Propose a change to the trajectory by randomly selecting whether
    to add a new kink (at a random imaginary time) or to remove an existing
    kink (selected in a random fashion). This is done according to the
    selection probabilities $g(N_{\!s}+1\rightarrow N_{\!s})$ and
    $g(N_{\!s}\rightarrow N_{\!s}+1)$.

    \item  Accept or reject the proposed change with probability
    $a_{\rm add}(N_{\!s}\rightarrow N_{\!s}+1)$ if
    adding, or $a_{\rm rem}(N_{\!s}+1\rightarrow N_{\!s})$ if removing
    a kink.

    \item  If the change has been accepted, then the {\it entire}
    trajectory that lies ``above'' the kink (i.e. from the imaginary time of the kink
    to $\beta$) is shifted across accordingly.
    If the proposed change has been rejected, then the trajectory is left untouched.
\end{enumerate}

\subsection{Analytical integration over kinks}
The Metropolis algorithm requires the action, which involves a
double integration in imaginary time, to be computed on each Monte
Carlo step. The fact that the trajectory consists of a series of
single kinks, between which the trajectory is a straight line
(${\bf r}(\bar{\tau})$ is constant), leads us to decompose the the
action $A$ in Eq.~\ref{GUlim} into segments according to
\begin{equation}
    A=\sum_{j=1}^{N_{\!s}\!+1} A_{jj}
    +2\sum_{j=1}^{N_{\!s}}\sum_{k=j+1}^{N_{\!s}\!+1} A_{jk}
    \label{CTAsum},
\end{equation}
where $j$ and $k$ label the kinks (along trajectories
corresponding with $\tau$ and $\tau'$ respectively in the double
integration), such that $\bar{\tau}_{j}$ is the imaginary-time at
which the $j$'th kink occurs, with $\bar{\tau}_{0}=0$ and
$\bar{\tau}_{N_{\!s}\!+1}=\bar\beta$.  We treat the diagonal $A_{jj}$
and off-diagonal segments $A_{jk}$ separately in order to increase
efficiency. Each $A_{jk}$ involves the range of imaginary-time
between successive kinks of
$\bar{\tau}_{j\!-\!1}\le\bar{\tau}\le\bar{\tau}_{j}$ and
$\bar{\tau}_{k\!-\!1}\le\bar{\tau}'\le\bar{\tau}_{k}$, in which
the electron coordinate is fixed at ${\bf r}(\bar{\tau})={\bf
r}(\bar{\tau}_{j\!-\!1})$ and ${\bf r}(\bar{\tau}')={\bf
r}(\bar{\tau}_{k\!-\!1})$ respectively. Thus the value of the
lattice summation $\Phi_{\Delta{\bf r}}\!\left({\bf
r}(\bar{\tau}_{j\!-\!1}),{\bf r}(\bar{\tau}_{k\!-\!1})\right)$
given by Eq.~\ref{CTphi2}, has a {\it constant} value, allowing
the double integration appearing in Eqs.~\ref{CTaper} and
\ref{CTdela} to be treated analytically for each segment $A_{jk}$.
The result after rearrangement reads
\begin{equation}
A = \frac{\lambda z}{\Phi_{0}^{(0,0)}} \Bigg\{
  \sum_{j=1}^{N_{s}+1}
   \Big[A_{0}^{(j)}\Phi_{0}^{(j,j)}
   +A_{1}^{(j)}\Phi_{\Delta{\bf r}}^{(j,j)} \Big]
+\sum_{j=1}^{N_{s}+1}\sum_{k=j+1}^{N_{s}+1} \Big[
   A_{2}\Phi_{0}^{(j,k)}
  +A_{3}\Phi_{\Delta{\bf r}}^{(j,k)}\Big] \Bigg\}
    \label{OBaction},
\end{equation}
where we have used the shorthand $\Phi_{\Delta{\bf
r}}^{(j,k)}=\Phi_{\Delta{\bf r}}\left({\bf
r}(\bar{\tau}_{j-1}),{\bf r}(\bar{\tau}_{k-1})\right)$
for the lattice summation defined in Eq.~\ref{CTphi2}, and
\begin{equation}
    A_{0}^{(j)}=\frac{1}{\bar{\omega}}\left[
     \bar{\omega}(\bar{\tau}_{j}-\bar{\tau}_{j-1})-K^{(j)}
     \right]
\label{OBa0},
\end{equation}
\begin{equation}
   A_{1}^{(j)}
   = \frac{1}{\bar{\omega}}e^{-\bar{\omega}(\bar{\beta}+\bar{\tau}_{j-1}-\bar{\tau}_{j})}
   \label{OBa1},
\end{equation}
\begin{equation}
   A_{2}
   = \frac{1}{\bar{\omega}}K^{(j)}K^{(k)}
   e^{-\bar{\omega}(\bar{\tau}_{k-1}-\bar{\tau}_{j})}
   \label{OBa2} ,
\end{equation}
and
\begin{equation}
   A_{3}
   = \frac{1}{\bar{\omega}}K^{(j)}K^{(k)}
   e^{-\bar{\omega}(\bar{\beta}-\bar{\tau}_{k}+\bar{\tau}_{j-1})}
     \label{OBa3},
\end{equation}
where we have defined
\begin{equation}
    K^{(j)}=
    1-e^{-\bar{\omega}(\bar{\tau}_{j}-\bar{\tau}_{j-1})}
    \label{k}.
\end{equation}
The action can thus be computed efficiently using this double
summation over kinks.

For the models studied in this paper, $f_{\bf m}(\bf n)$ depends
only on the {\it relative} lattice distance $|{\bf m}-{\bf n}|$,
and tends to zero (or is zero) at large distance. Consequently,
the lattice summation, Eq.~\ref{CTphi2}, is a function of the {\it
single} variable ${\bf r}'={\bf r}_{2}-{\bf r}_{1}-\Delta{\bf r}$
only, namely
\begin{equation}
    \Phi({\bf r}')=
    \sum_{{\bf m}'} \bar{f}({\bf m}')\bar{f}({\bf m}'-{\bf r}')
    \label{SDphi2},
\end{equation}
which can be evaluated for all possible values of ${\bf r}'$ in
advance of the simulation proper, improving the efficiency of the
QMC scheme.

\subsection{Physical Observables}   \label{obs}
We consider four physical observables:  the ground state energy,
the number of phonons in the polaron cloud, the effective mass,
and the isotope exponent on the effective mass. For a given
observable $Q$, the expectation value is the statistical average
over trajectories at ${\bf P}=0$ (ground state), which can be
written in the form
\begin{equation}
        \langle Q \rangle_{0}
    =\frac{ \int_{\rm tw}\mathcal{D}{\bf r} \ Q w(N_{s})}
          { \int_{\rm tw}\mathcal{D}{\bf r} \ w(N_{s})}
    \label{CQexpo},
\end{equation}
where the phonon degrees of freedom have been integrated out, and
$w(N_{s})$ is given by Eq.~\ref{CQwsp}.  This corresponds
to a simple arithmetic average within the QMC simulation.

The ground state (${\bf P}=0$) energy estimator is given by
\cite{Kornil}
\begin{equation}
    E_{0}(0)= -\left\langle \frac{1}{w}\frac{\partial w}{\partial\beta}
    \right\rangle_{\!\! 0}
     = - \left\langle \frac{\partial A}{\partial\beta}\right\rangle_{\!\! 0}
     -\frac{1}{\beta} \left\langle \sum_{s}N_{s}\right\rangle_{\!\! 0}
    \label{CQtote},
\end{equation}
which follows from the corresponding finite-imaginary-time energy
estimator \cite{Raedt} in the continuum limit. Within the QMC
simulation, then, we must gather separate statistics for the total
number of kinks $\sum_{s}N_{s}$ and the quantity $\partial
A/\partial\beta$. One can see that the expression for $\partial
A/\partial\beta$ is easily obtained by analytically
differentiating Eq.~\ref{OBaction} with respect to $\beta$.

The number of phonons in the polaron cloud $N_{\rm ph}$ quantifies
the amount of lattice deformation associated with the polaron. The
value of $N_{\rm ph}$ is given by the expectation value of the
phonon number operator, which can be isolated from the model
Hamiltonian using the fact that $ \left. \partial
H/\partial(\hbar\omega)\right|_{\lambda\omega} = \sum_{\bf m}
d_{\bf m}^{\dagger} d_{\bf m}$. This can be related to the action
via the free energy $F_{0}= -{\beta}^{-1}\ln Z_{0}$ to give
\begin{equation}
    N_{\mathrm{ph}}  =   \left\langle \sum_{\bf m} d^{\dagger}_{\bf m} d_{\bf m}
     \right\rangle_{\!\! 0}
    = \left. \frac{\partial
F_{0}}{\partial(\hbar\omega)}\right|_{\lambda\omega}
    = -\frac{1}{\bar{\beta}}
    \left\langle \left. \frac{\partial
    A}{\partial\bar{\omega}}\right|_{\lambda\bar{\omega}}
     \right\rangle_{\!\! 0}
    \label{OBnph},
\end{equation}
where $\left.\partial A/
\partial\bar{\omega}\right|_{\lambda\bar{\omega}}$ is easily
obtained by differentiating Eq.~\ref{OBaction} with respect to
$\bar{\omega}$ holding the product $\lambda\bar{\omega}$ constant.

As discussed in section (\ref{secteff}), by imposing twisted
boundary conditions in imaginary time, dynamical properties can be
inferred directly from QMC simulation. The effective mass of the
polaron $m^{*}$, for the isotropic or one-dimensional case, may be
measured using
\begin{equation}
    \frac{m_{0}}{m^{*}} = \frac{1}{2\bar{\beta}}
    \left\langle\left( \Delta {\bf r}\right)^{2}\right\rangle_{\!\! 0}
    \label{expextm},
\end{equation}
where the difference in position of the endpoints of the
trajectory $\Delta{\bf r}= {\bf r}(\bar{\beta}) - {\bf r}(0)$ is
measured in units of the lattice constant $a$, and $m_{0}$ is the
bare electron mass.

The isotope effect is most often observed via its influence on the
superconducting transition temperature $T_{\rm c}$.  The
dependence of $T_{\rm c}$ on the mass of the lattice ions $M$ has
been found empirically to be $T_{\rm c}\propto M^{-\alpha}$~\cite{Max,Reyn}, where $\alpha$ is known as the isotope exponent on $T_{c}$. In a similar way, let
us define the {\it isotope exponent on the effective mass}
$\alpha_{m^{*}}$ as
\begin{equation}
    \alpha_{m^{*}} = \frac{M}{m^{*}}\frac{dm^{*}}{dM}
     = -M \frac{m^{*}}{m_{0}}\frac{\partial}{\partial M}
    \left( \frac{m_{0}}{m^{*}} \right)
    \label{ap1}
\end{equation}
for the isotropic or one-dimensional case. On substitution of the
derivative of $m_{0}/m^{*}$, as in Eq.~\ref{expextm}, with respect
to $M$, we have
\begin{equation}
    \alpha_{m^{*}} = \frac{\bar{\omega}}{2}
    \frac{1}{\left\langle\left( \Delta{\bf r}\right)^{2}\right\rangle_{\!\! 0}}
    \left[
\left\langle\left( \Delta{\bf r}\right)^{2}\frac{\partial
A}{\partial \bar{\omega}} \right\rangle_{\!\! 0}
  -\left\langle\left( \Delta{\bf r}\right)^{2}\right\rangle_{\!\! 0}
   \left\langle\frac{\partial A}{\partial \bar{\omega}} \right\rangle_{\!\! 0}
    \right]
    \label{iso2},
\end{equation}
where the ionic mass enters our formalism via the phonon frequency
$\omega = ({k/M})^{1/2}$, where $k$ is some ``spring constant'',
and $\partial A/\partial\bar{\omega}$ is easily obtained from
Eq.~\ref{OBaction}.  (For a general $D$-dimensional system, we may
define the $d$'th component of the isotope exponent on the effective
mass as $\alpha^{(d)}_{m^{*}} = -M(m_{d}^{*}/m_{0,d})\partial
/\partial M \left( m_{0,d}/m_{d}^{*} \right)$, where
$m_{0,d}/m_{d}^{*} = (2\bar{\beta})^{-1}\left\langle\left( \Delta
r_{d}\right)^{2}\right\rangle_{0}$, with $\Delta r_{d} =
r_{d}(\bar{\beta}) - r_{d}(0)$, and thus we may write
$\alpha^{(d)}_{m^{*}}$ as Eq.~\ref{iso2} with every $\Delta{\bf
r}$ replaced by $\Delta r_{d}$).

\subsection{Simulation details}
The QMC scheme is based on the simulation of a {\it single}
trajectory ${\bf r}(\bar{\tau})$ in imaginary time
$0\le\bar{\tau}\le\bar\beta$. The standard Metropolis algorithm is
used to alter the shape of the trajectory by the addition and
deletion of single kinks, as described in section (\ref{algo}).
The start of the trajectory ${\bf r}(0)$ does not change
throughout the simulation, but the other end ${\bf
r}(\bar{\beta})$ is ``free'' (open boundary conditions in
imaginary time). In practice, the shape of the trajectory was
represented using a list containing the imaginary-time, and the
direction, of each kink. In addition, we also kept track of the
value of $\Delta {\bf r}={\bf r}(\bar{\beta}) -{\bf r}(0)$, and
the the total number of kinks of each direction $\{N_{s}\}$. The
major computational task is the evaluation, on each Monte Carlo
step, of the action given by Eq.~(\ref{OBaction}).
(The number of exponential-function-evaluations was reduced by
storing the values of $A_{1}^{(j)}$ and $K^{(j)}$ along with each
kink, reducing the overall computational effort). In order to
calculate the expectation values of the observables given in
section (\ref{obs}), separate statistics for the quantities
$\langle\sum_{s}N_{s}\rangle_{0}$, $\langle\partial
A/\partial\bar{\beta}\rangle_{0}$, $\langle\left.
\partial A/\partial\bar{\omega}\right|_{\lambda\bar{\omega}}\rangle_{0}$,
$\langle\left( \Delta {\bf r}\right)^{2}\rangle_{0}$,
$\langle\partial A/\partial\bar{\omega}\rangle_{0}$, and
$\langle\left( \Delta{\bf r}\right)^{2}\partial
A/\partial\bar{\omega}\rangle_{0}$ were gathered every 10-50 Monte
Carlo steps. We only consider the case of ${\bf P}=0$,
corresponding to the ground state of the system, where there is no
sign problem.

The four one-dimensional interaction models studied differ only in
the value of the screening length $R_{\rm sc}$. The model
dependency enters the simulation via $\Phi({\bf r}')$ given by
Eq.~\ref{SDphi2}. For each model, simulations were conducted for
various different values of the dimensionless parameters
$\bar{\omega}$ and $\lambda$.

The value of $\bar{\beta}$ was set at a sufficiently large value
to enforce the low temperature limit
$\exp(\bar{\omega}\bar{\beta})\to\infty$.  For the present
simulations, a value of $\bar{\beta}\ge 15$ was found to make the
finite-temperature error negligible. (Reducing the value of
$\bar{\omega}$, or increasing $\lambda$, beyond those studied here
would require this value of $\bar{\beta}$ to be increased).
Increasing the value of $\bar{\beta}$ increases the ``length'' of
the trajectory (which involves more kinks), and thus increases the
computational effort required to perform each Monte Carlo step.

For each set of model parameters ($R_{\rm sc}$, $\bar{\omega}$ and
$\lambda$), between 3 and 6 statistically independent Monte Carlo
runs were performed, each using a different value of the inverse
temperature $15\leq\bar{\beta}\leq 25$.  The number of Monte Carlo
steps in each run was taken to be about five times the ``warm-up''
period. Typically, the runs consisted of between $1\times 10^{7}$
and $5\times 10^{7}$ steps in total.  (The statistics gathered for
each set of runs were viewed together graphically, in order to
better estimate the point at which equilibrium had been reached).
Only those statistics gathered after the estimated warm-up period
were included in the averages. Given that the finite-temperature
error is small, and in the absence of systematic finite-size and
finite-time-step errors, the main source of error is statistical.
The size of the statistical error depends on $f_{\bf m}({\bf n})$
and $\bar{\beta}$, as well as on $\bar{\omega}$ and $\lambda$. For
each set of model parameters ($R_{\rm sc}$, $\bar{\omega}$ and
$\lambda$), we performed a sufficient number of runs to ensure
that the Monte Carlo averages were determined to a statistical
error of less than one percent.

\section{Limiting cases}               \label{sect5}   
\subsection{Strong Coupling (Small Polaron) Regime}\label{sect5a}
When the electron-phonon coupling is strong, the electron becomes
``trapped'' in a potential well created by the induced lattice
distortion.  In this case the ``size'' of the polaron state can
become comparable with the lattice constant, and the term ``small
polaron'' is used. The condition for small polaron formation is
$\lambda=E_{\rm p}/zt\geq1$, which is referred to as the
strong-coupling regime \cite{Alex0}. The small polaron can move
from site to site (at zero temperature) through the action of
zero-point motion.

An analytical method to determine the effective mass and energy
dispersion in the strong-coupling regime, for a lattice polaron
with a general electron-phonon interaction force  $f_{\bf m}({\bf
n})$\cite{Korn2}, is based on the Lang-Firsov canonical
transformation \cite{fir} (which renders the transformed
Hamiltonian diagonal for $\lambda\to\infty$), followed by a
second-order perturbation technique that uses $1/\lambda$ as a
small  parameter \cite{Alex0}.  For nearest-neighbor hopping
only, and forms of $f_{\bf m}({\bf n})$ that depend on the
relative lattice distance $|{\bf m}-{\bf n}|$, the result for  the
lowest energy levels reads \cite{Korn2}
\begin{equation}
E({\bf k})=-E_{\rm p} -  \varepsilon_{\rm p}({\bf k}) - \sum_{{\bf
k'},\{n_{\bf q}\}} \frac{ \left| \langle{\bf k},0\mid \sum_{\bf
nn'}t_{\bf nn'} \exp\left[\sum_{\mathbf{mq}}\frac{(f_\mathbf{m}(\mathbf{n})-f_\mathbf{m}(\mathbf{n}^\prime))(e^{i\mathbf{q\cdot m}}d_{\mathbf q}^\dag-e^{-i\mathbf{q\cdot m}}d_{\mathbf q})}{\sqrt{2NM\hbar\omega_{\mathbf q}^3}}\right]c_{\bf n'}^{\dagger}c_{\bf n} \mid{\bf
k'},\{n_{\bf q}\}\rangle\right|^{2}} {\hbar\omega\sum_{\bf q}n_{\bf q}}
\label{Ek},
\end{equation}
where the summation is over intermediate states with one or more phonons, $\sum_{\bf q}n_{\bf q}>0$, the polaronic  energy-level shift
\begin{equation}
        E_{\rm p}=\frac{1}{2M\omega^{2}}\sum_{{\bf m}} f_{\bf
    m}^{2}(0) \label{Eep}
\end{equation}
(used to define $\lambda$) corresponds with the solution for
$\lambda\to\infty$, and the small-polaron dispersion
\begin{equation}
    \varepsilon_{\rm p}({\bf k})= t\sum_{{\bf n}\neq 0}
    e^{-g^{2}(1)}e^{-i{\bf k}\cdot{\bf n}} \label{SCpdis2},
\end{equation}
where $g^{2}$ is  known as the (zero temperature)  {\it mass
renormalization exponent}, given by
\begin{equation}
     g^{2}(1)=\frac{1}{2M\hbar\omega^{3}} \sum_{{\bf m}}\left[
 f_{\bf m}^{2}(0) -  f_{\bf m}(0)f_{{\bf m}}(1)\right]
 \label{Eg2}.
\end{equation}
The final term of Eq.~\ref{Ek} is quadratic with respect to
the bare hopping amplitude $t_{{\bf nn}'}$, and  produces a
negative correction  to the energy. (Here, $n_{\bf q}$ is the
number of phonons with wave vector ${\bf q}$, and  $ \mid{\bf
k'},n_{\bf q}\rangle$ is an excited state that consists of a
single electron with wave vector ${\bf k'}$ and one or more
phonons.)  It is of order of $1/\lambda^{2}$ and is almost {\it
independent} of ${\bf k}$ \cite{Korn2}.

In the ground state of the system (${\bf k}=0$), the value of
$\varepsilon_{\rm p}({\bf k})$ is real and small compared with
$-E_{\rm  p}$, and so Eq.~\ref{Ek} is dominated by $-E_{\rm
P}$. Thus, the (dimensionless) strong-coupling result for the
ground state energy is
\begin{equation}
     \frac{E_{0}}{t}=z\lambda \label{edim},
\end{equation}
and since $E_{\rm p}=N_{\rm ph}\hbar\omega$, it follows that the
number of phonons in the polaron cloud at ${\bf k}=0$ is given by
\begin{equation}
    N_{\rm{ph}} =\frac{z\lambda}{\bar{\omega}}  \label{ndim}.
\end{equation}
Assuming that the third term in Eq.~\ref{Ek} is {\it
completely} independent of $\mathbf k$, then the inverse effective mass,
for the isotropic or one-dimensional case, is given by
\begin{equation}
    \frac{m_{0}}{m^{*}}=m_{0}\left. \frac{\partial^{2}E(k)}{\partial
    k^{2}}\right|_{k\to 0}=e^{-g^{2}(1)} \label{EMmom*}.
\end{equation}
This can be conveniently expressed in dimensionless form as
\begin{equation}
    \frac{m_{0}}{m^{*}}=\exp\left(\frac{-2\gamma
    \lambda}{\bar{\omega}}\right)   \label{SCmassdim},
\end{equation}
where we have defined the constant
\begin{equation}
\gamma\equiv 1-\frac{\sum_{\bf m}f_{\bf m}(0)f_{\bf m}(1)}
{\sum_{\bf m^\prime}f_{\bf m^\prime}^{2}(0)}, \label{gamma}
\end{equation}
which depends only on the {\it shape} of the electron-phonon
interaction force. The isotope exponent on the effective mass
$\alpha_{m^{*}}$, defined in Eq.~\ref{ap1}, can be written
in terms of $\omega \propto M^{-1/2}$ as
\begin{equation}
    \alpha_{m^{*}}=\frac{m^{*}}{m_{0}}
    \frac{\omega}{2}\frac{\partial}{\partial\omega} \left(
    \frac{m_{0}}{m^{*}}\right) \label{WCisodef2},
\end{equation}
for the isotropic (or one-dimensional) case.  Thus, the
(dimensionless) strong-coupling isotope exponent is given by
\cite{ale3}
\begin{equation}
    \alpha_{m^{*}} = \frac{\gamma\lambda}{\bar{\omega}} \label{EMal}.
\end{equation}
The value of the dimensionless constant $0\leq\gamma\leq 1$ must
in general be determined numerically: the value calculated for
each of the one-dimensional interaction models studied in this
work is presented in TABLE \ref{table1}.
\begin{table}
    \centering
\begin{tabular}{|c|c|c|} \hline
    Interaction Model & $\gamma$ \\ \hline Holstein & 1.000 \\
    Screened Fr\"{o}hlich ($R_{sc}=1$) & 0.745 \\ Screened
    Fr\"{o}hlich ($R_{sc}=3$) & 0.531 \\ Non-screened Fr\"{o}hlich
    & 0.387 \\
    \hline \end{tabular}
 \caption{Values of the dimensionless parameter
    $\gamma=1-\sum_{\bf  m}f_{\bf m}(0)f_{\bf m}(1)/\sum_{\bf m'}
    f_{\bf m'}^{2}(0)$ for  the one-dimensional models studied in
    this work, where  $\gamma$ depends only on the {\it shape} of
    the interaction force.}
 \label{table1}
\end{table}

\subsection{Weak Coupling (Large Polaron) Regime}\label{sect5b}
In 1950 Fr\"{o}hlich \emph{et al.} \cite{Froh} considered the
ground state of a polaron in the weak-coupling limit $\lambda\ll
1$, using second-order perturbation theory, under the condition
that there is never more than one phonon virtually excited, and
the discreteness of the lattice is unimportant (because the
band-electron-like ``large polaron'' state \cite{pek} is much
larger than the lattice constant).  In our case we use the tight-binding dispersion
\begin{equation}
    \varepsilon_{0}({\bf k})=  - 2t\cos({k}a)
\label{WCband}
\end{equation}
rather than the parabolic approximation
$\varepsilon_{0}({\bf k})= \hbar^2{\bf k}^{2}/2m_{0} + O({{\bf
k}^{4}})$, where $m_0 = \hbar^2/(2ta^2)$ used by Fr\"{o}hlich\cite{Froh}.  Thus we write
\begin{equation}
    H=\varepsilon_{0}({\bf k}) +
\hbar\omega\sum_{\bf q}d_{\bf q}^{\dag}d_{\bf q} + H_{\rm e-ph}
\label{WCpert},
\end{equation}
where the electron-phonon term $H_{\rm e-ph}$ is a small
perturbation.  Assuming that ${f}_{\bf
m}({\bf n})$ depends only on the relative lattice distance $|{\bf
m}-{\bf n}|$, then $H_{\rm e-ph}$ given by Eq.~\ref{MHheph} may be
written in momentum representation as
\begin{equation}
    H_{\rm e-ph}=-\sqrt{\tilde\kappa}\sum_{\bf q,k}\tilde{f}_{\bf q}\left(
    c_{\bf  k-q}^{\dag}c_{\bf  k}d_{\bf q}^{\dag}+{\rm
    h.c.}\right) \label{WCpert1},
\end{equation}
where we have defined
\begin{equation}
    \tilde\kappa=\frac{z\lambda\bar{\omega}t^2}
    {\sum_{\bf m}\bar{f}_{\bf m}^{2}(0)}
    \label{kappa},
\end{equation} and
\begin{equation}
    \tilde{f}_{\bf q} = \sum_{\bf r}\bar{f}_{\bf r}(0)e^{-i{\bf
    q\cdot r}/\hbar} \label{WCftran}.
\end{equation}
Here, the summation is over all values of ${\bf  r}={\bf m}-{\bf
n}$ for which $\bar{f}_{\bf m}({\bf n})=\bar{f}({\bf m}-{\bf
n})=\bar{f}_{\bf r}(0)$ is a function of a single variable only.
Using standard second order perturbation theory, with an initial
state that consists of an electron of momentum $\hbar{\bf k}$ and
no phonons, and an intermediate state that consists of an electron
with momentum $\hbar({\bf k-q})$ and a single phonon of momentum
$\hbar{\bf q}$, the energy measured from the bottom of the
electron band is given by
\begin{equation}
    E({\bf k})= \varepsilon_{0}({\bf k})
    -\tilde\kappa\sum_{{\bf q}}
     \frac{\mid  \tilde{f}_{\bf q}\mid^{2}}{W}
     \label{WCEE},
\end{equation}
where we have defined
\begin{equation}
    W=\varepsilon_{0}({\bf k}-{\bf q})+\hbar\omega - \varepsilon_{0}({\bf k})
    \label{W}.
\end{equation}
Here the ground state energy $E_{0}$ occurs at ${\bf k}={\bf 0}$
and $\sum_{\bf q}$ is a suitable Brillouin zone average.

The number of virtual phonons in the polaron cloud is defined as
$N_{\rm ph}  =\sum_{\bf q}\langle 0'\mid{d_{\bf q}^{\dag}d_{\bf
q}}\mid 0'\rangle$ where $\mid 0' \rangle$ represents the
eigenstate for the perturbed Hamiltonian.  Using standard first
order perturbation theory, this is given by
\begin{equation}
    N_{\rm ph}=\tilde\kappa \sum_{\bf
    q}\frac{|\tilde{f}_{\bf q}|^{2}}{W^{2}}
     \label{WCnph3a}.
\end{equation}
The effective mass is easily found by differentiating
Eq.~\ref{WCEE} twice with respect to ${\bf k}$, according to
Eq.~\ref{EMmom*}, to give
\begin{equation}
    \frac{m_{0}}{m^{*}}=
    1-\frac{\tilde\kappa}{2at^{2}}\sum_{\bf q}
    \frac{\mid  \tilde{f}_{\bf q}\mid^{2}
    \left( 2W'^{2} - W''W \right)}{W^{3}}
    \label{WCm2},
\end{equation}
where $W'=\partial W/\partial {\bf k}$ and $W''=\partial^{2}
W/\partial {\bf k}^{2}$. Differentiating this expression with
respect to $\omega$, according to equation (\ref{WCisodef2}),
gives the isotope exponent on the effective mass as
\begin{equation}
\alpha_{m^{*}} =  \frac{\tilde\kappa}{4at^{2}}
\frac{m^{*}}{m_{0}}\sum_{\bf q} \left[
    \frac{\mid  \tilde{f}_{\bf q}\mid^{2} W''(W-2\hbar\omega)}{W^{3}}
    -\frac{\mid  \tilde{f}_{\bf q}\mid^{2} 2W'^{2}(W-3\hbar\omega)}{W^{4}}
    \right] \label{WCalph1}.
\end{equation}

The above expressions for the observables may be written in the
form
\begin{equation}
    \frac{1}{t}E_{ 0}  =  -2-\lambda
    \Upsilon_{E_{0}}(\bar{f},\bar{\omega}) \label{WCNe2},
\end{equation}
\begin{equation}
    N_{\rm ph} = \lambda \Upsilon_{N_{\rm
    ph}}(\bar{f},\bar{\omega}) \label{WCNn},
\end{equation}
\begin{equation}
    \frac{m_{0}}{m^{*}}=
    1-\lambda\Upsilon_{m^{*}}(\bar{f},\bar{\omega}) \label{WCNm},
\end{equation}
and
\begin{equation}
\alpha_{m^{*}} = \frac{\lambda}{m_{0}/m^{*}}
                \Upsilon_{\alpha_{m^{*}}}(\bar{f},\bar{\omega})
    \label{WCNal},
\end{equation}
where $\Upsilon(\bar{f},\bar{\omega})$ is the coefficient of the
linear term in $\lambda$, which may be calculated numerically. We
are interested in the ground state properties and thus must
evaluate the above expressions at ${\bf k}\rightarrow 0$. The
ground state values of $\Upsilon(\bar{f},\bar{\omega})$ for each
one-dimensional interaction model are presented in TABLE
\ref{numval} for $\bar{\omega}=1$.

\begin{table}[h!]
    \centering \begin{tabular}{|c|c|c|c|c|c|}
    \hline Interaction Model & $\sum_{\bf m}\bar{f}_{\bf
    m}^{2}(0)$  & $\Upsilon_{E_{ 0}}$ & $\Upsilon_{N_{\rm ph}}$ &
    $\Upsilon_{{m_{0}}/{m^{*}}}$ & $\Upsilon_{\alpha_{m^{*}}}$ \\
    \hline Holstein ($R_{\rm sc}\to 0$)& 1.000 & 0.894 & 0.537 &
    0.537 & 0.125\\
Screened Fr\"{o}hlich ($R_{\rm sc}=1$) & 1.034
    & 1.080 & 0.736 &0.625 & 0.200 \\
Screened Fr\"{o}hlich
    ($R_{\rm sc}=3$) & 1.133 & 1.257 & 0.938 & 0.678 &  0.266 \\
    Unscreened Fr\"{o}hlich ($R_{\rm sc}\to\infty$)& 1.269 &
    1.394 & 1.104 & 0.687 & 0.306 \\
    \hline \end{tabular}
  \caption{The numerically calculated values of the coefficient of the linear term in  $\lambda$
   in Eqs.~\ref{WCNe2}--\ref{WCNal}, for each one-dimensional interaction model at
   $\bar{\omega}=1$ in the weak coupling limit.
   For example, the ground state energy (\ref{WCNe2}) of the Holstein polaron is given by
    $E_{0}/t=-2-0.894\lambda$. }
  \label{numval}
\end{table}

The coefficients can be expressed in closed form for the simplest
case of the one-dimensional Holstein interaction. Here we find
that the energy is

\begin{equation}
E({\bf k})  =  -2t\cos(ka)- \frac{2\bar\omega t\lambda} {\left[( \bar\omega + 2\cos (ka) )^2 - 4 \right]^{1/2}}\\
,
\end{equation}
so that
\begin{equation}
\Upsilon_{E_{0}}(\bar{f},\bar{\omega})  =   \frac{2\bar\omega} {\left( \bar\omega^2 + 4\bar\omega \right)^{1/2}} \label{tb1}\\
,
\end{equation}
\begin{equation}
\Upsilon_{N_{\rm ph}}(\bar{f},\bar{\omega})  =   \frac{2\bar\omega(\bar\omega+2)} {\left( \bar\omega^2 + 4\bar\omega \right)^{3/2}}  \label{tb2}\\
,
\end{equation}
\begin{equation}
\Upsilon_{m^{*}}(\bar{f},\bar{\omega})  =  \frac{2\bar\omega(\bar\omega+2)} {\left( \bar\omega^2 + 4\bar\omega \right)^{3/2}}  \label{tb3}\\
,
\end{equation}
and
\begin{equation}
\Upsilon_{\alpha_{m^{*}}}(\bar{f},\bar{\omega})  =
\frac{\bar\omega^2(\bar\omega^2+2\bar\omega+4)} {\left(
\bar\omega^2 + 4\bar\omega \right)^{5/2}}  \label{tb4}.
\end{equation}

\section{Results}\label{sect6}   
\subsection{Holstein Interaction}
\subsubsection{Self-Trapping Transition}
\begin{figure}
\includegraphics[scale=0.7, viewport=0 400 800 800,clip]{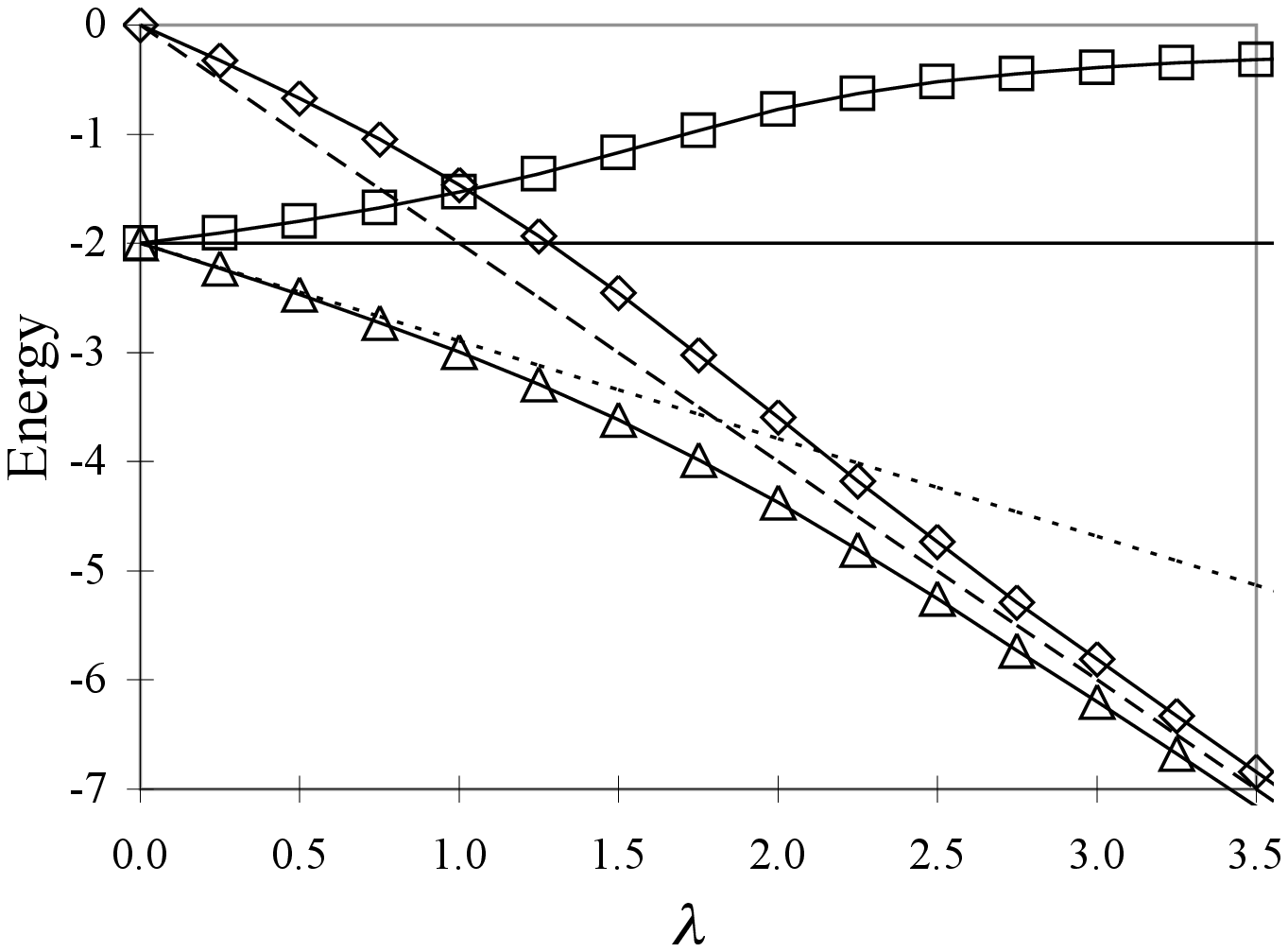}
\centering \caption{The variation of ground state energy
$E_{0}(0)$ (triangles), together with potential energy (diamonds) and kinetic energy
(squares), with coupling $\lambda$ for the one-dimensional Holstein
model with a dimensionless phonon frequency of $\bar{\omega}=1$. The dashed line
is the strong coupling perturbation (SCP) result (\ref{edim}) and the dotted
line the weak-coupling perturbation (WCP) result (\ref{WCNe2}). One can clearly
distinguish the large-polaron, transition and small-polaron
regions.} \label{E}
\end{figure}

\begin{figure}
\includegraphics[scale=0.5, angle=-90]{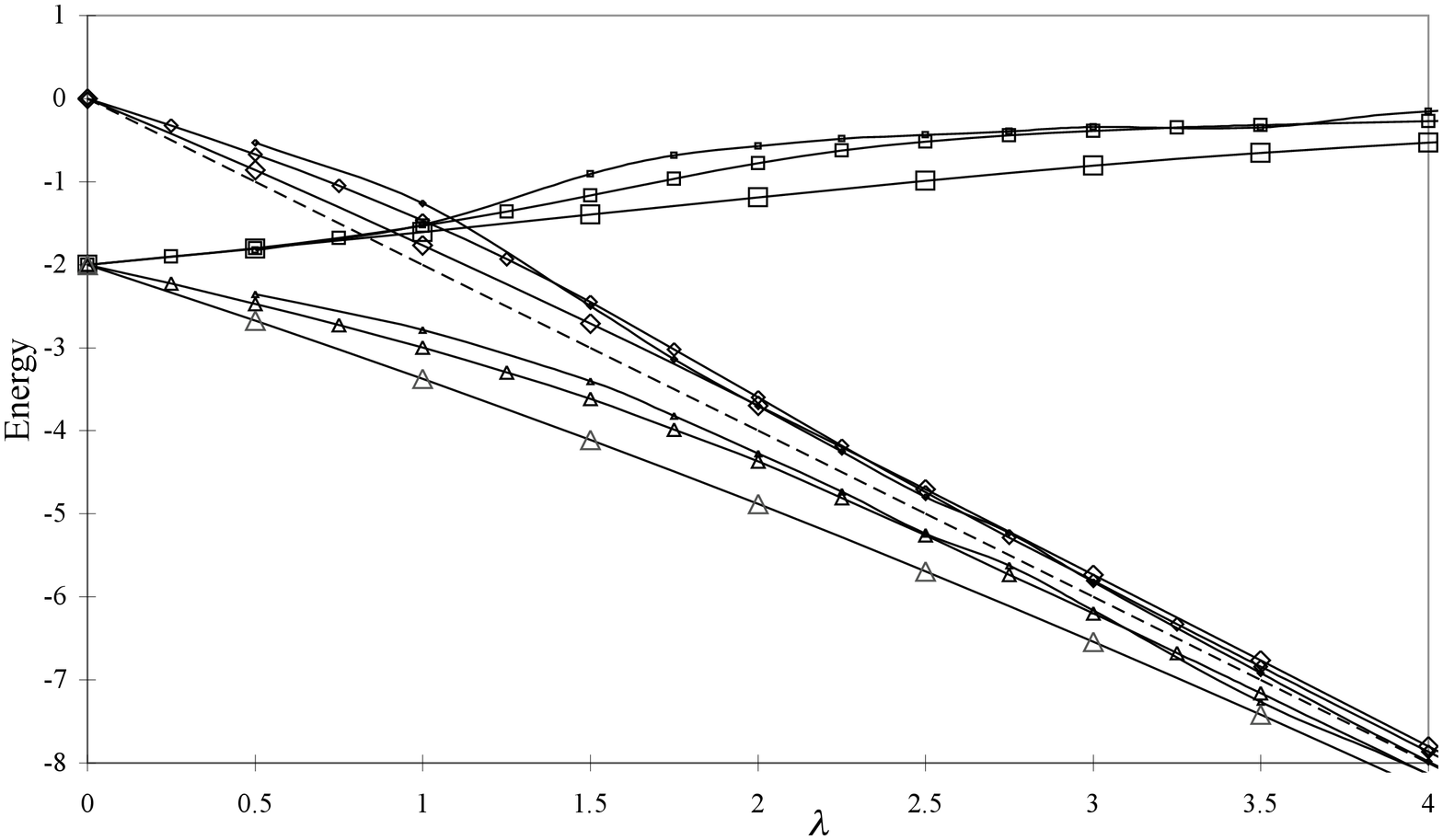}
    \caption{The variation of ground state energy
$E_{0}(0)$ (triangles), together with potential energy (diamonds) and kinetic energy
(squares), with coupling $\lambda$, for the one-dimensional Holstein model with a phonon frequency of
$\bar{\omega}=0.5$ (small symbols), $1.0$ (medium-size symbols) and $3.0$ (large symbols). As $\bar{\omega}$ increases,
the transition region shifts to higher $\lambda$ and becomes
broader.}
    \label{Eall}
\end{figure}
\begin{figure}
\includegraphics[scale=0.5, angle=-90]{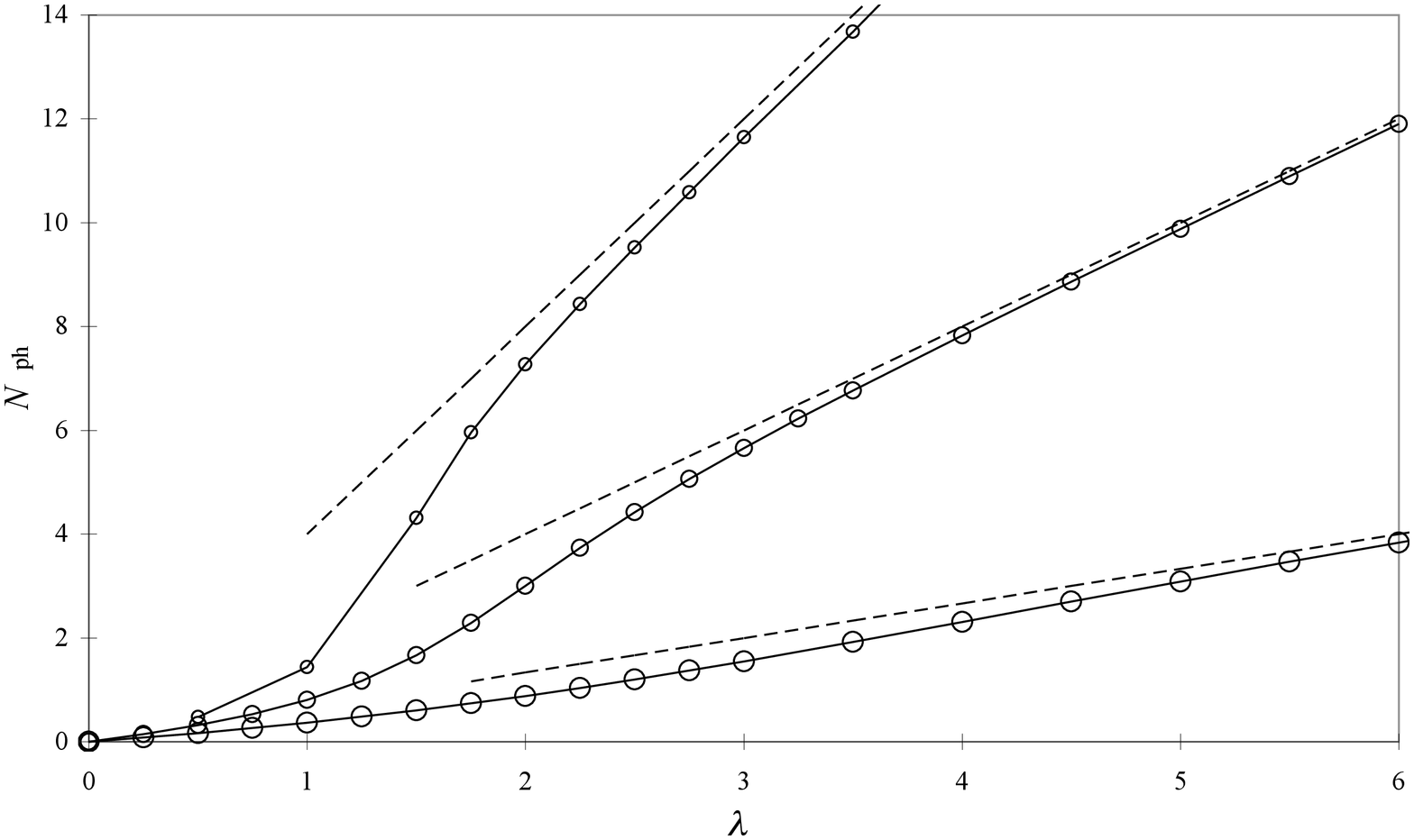}
    \caption{The number of phonons for the Holstein model at
    $\bar{\omega}=0.5, 1.0, 3.0$ (small, medium-size, and large circles respectively). The dashed line shows the SCP
    result (\ref{edim}). The transition region
    boundaries are the same as those in FIG.~ \ref{Eall}.}
    \label{Nph}
\end{figure}

\begin{figure}
\includegraphics[scale=0.5, viewport=0 350 800 850,clip]{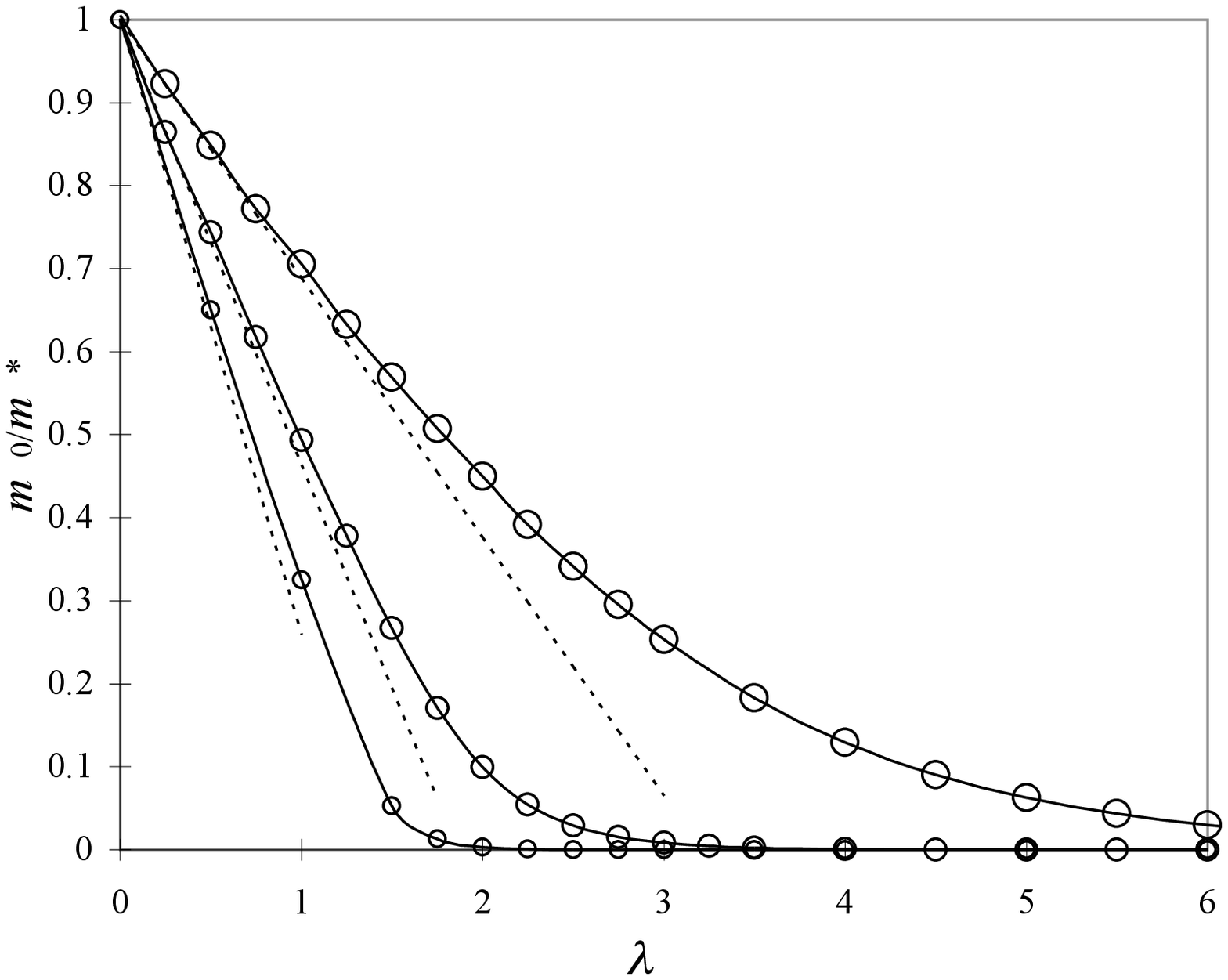}
    \caption{The inverse effective mass $m_{0}/m^{*}$
    (where $m_{0}=\hbar^{2}/2ta^{2}$ is the bare-electron mass)
    for the Holstein model at $\bar{\omega}=0.5, 1.0, 3.0$ (small, medium-size, and large circles respectively).
    The dotted line shows the WCP result (\ref{tb3}) for the tight-binding Hamiltonian.}
    \label{minv}
\end{figure}
\begin{figure}
\includegraphics[scale=0.5, angle=-90]{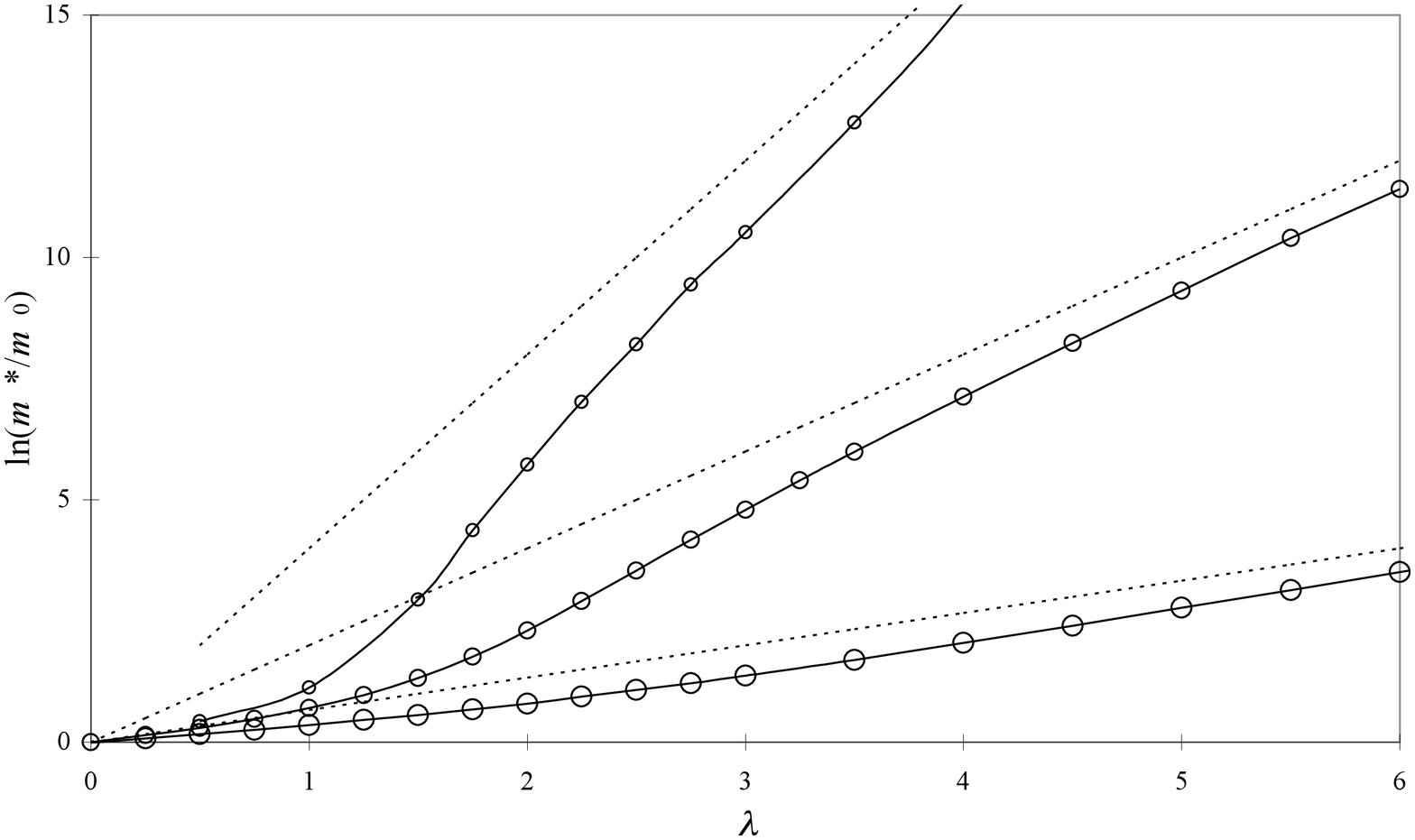}
    \caption{The logarithm of the effective mass for the Holstein model at
    $\bar{\omega}=0.5, 1.0$ and $3.0$ (small, medium-size, and large circles respectively). The dashed line shows the SCP
    result $\ln(m^{*}/m_{0})=2\lambda/\bar{\omega}$.}
    \label{m}
\end{figure}

\begin{figure}
\includegraphics[scale=0.5,angle=-90]{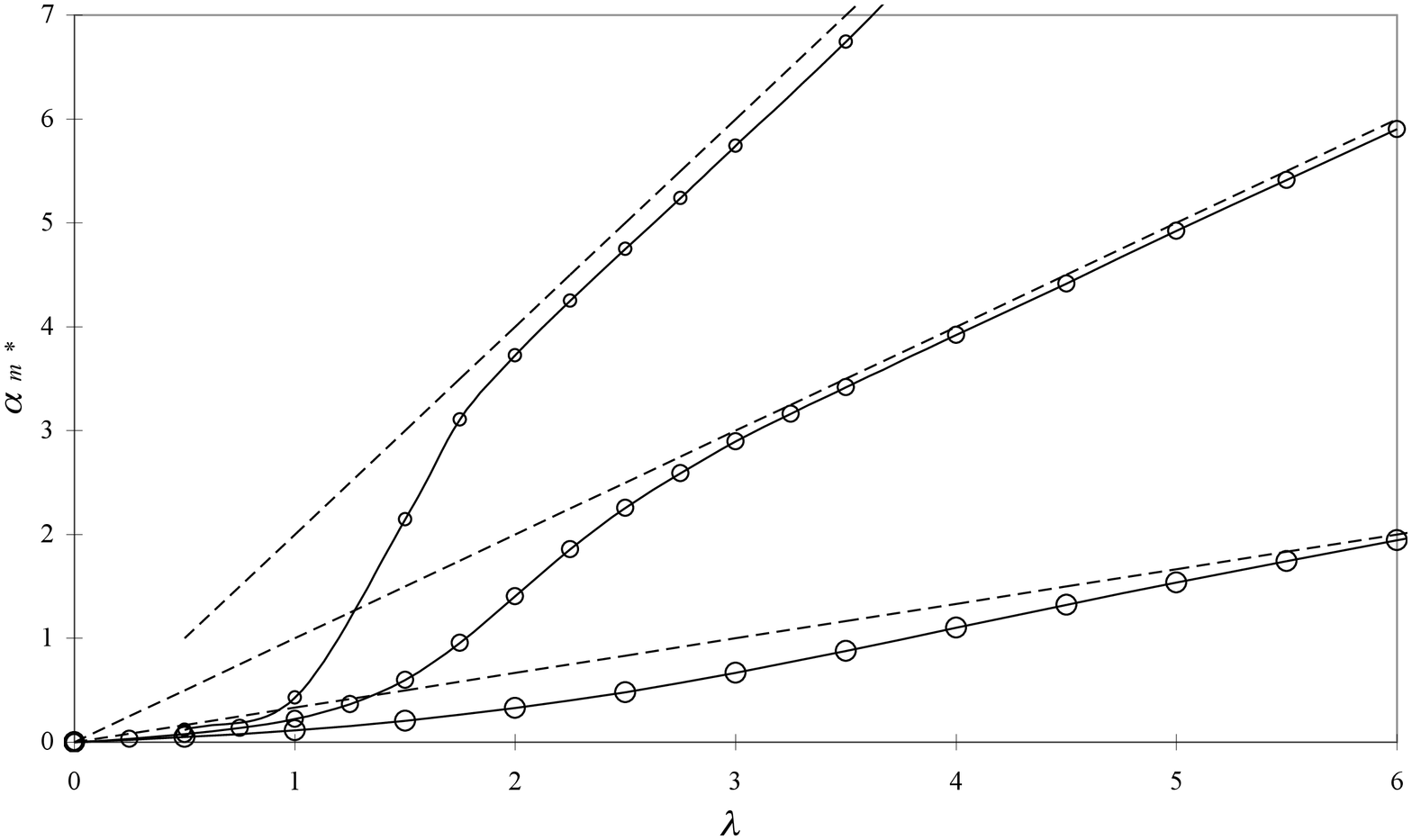}
    \caption{The isotope exponent for the Holstein model at
    $\bar{\omega}=0.5, 1.0, 3.0$ (small, medium-size, and large circles respectively). The dotted line shows the SCP
    result.}
    \label{iso}
\end{figure}

Let us consider first the QMC results for the simplest case of the
one-dimensional Holstein model. The variation of the ground state
energy $E_{0}(0)$ with the coupling constant $\lambda$ is shown in
FIG.~ \ref{E} for a fixed phonon frequency of $\bar{\omega}=1$.
The $E_{0}(0)$ curve tends to the weak-coupling perturbation (WCP)
result from below as $\lambda\rightarrow 0$, and to the
strong-coupling perturbation (SCP) result as
$\lambda\rightarrow\infty$. Here we have also plotted the first
and second terms of Eq.~\ref{CQtote}, which provide an
indication of the potential energy (PE) and the kinetic energy
(KE) of the system, respectively. Three separate regions can be
clearly distinguished from FIG.~ \ref{E}:
\begin{enumerate}
\item The {\it large-polaron region} at weak-coupling ($\lambda=0$
to $\lambda\approx 1$ in this case) is defined as the range of
$\lambda$ for which $E_{0}(0)$ is accurately described by WCP
theory (section \ref{sect5b}).  At $\lambda =0$ the energy is
entirely kinetic and represents the bottom of the bare-electron
band (with $E_{0}(0)=-zt$).  As $\lambda$ increases within this
region, KE increases but remains large compared with PE
(non-localized band-like states). \item The {\it
small-polaron region} at strong-coupling (in this case $\lambda
\approx 2.5$ to $\lambda \rightarrow\infty$) has $E_{0}(0)$
accurately predicted by SCP theory (section \ref{sect5a}). The PE
is much greater than the KE (self-trapped states).

\item The smooth {\it transition region} at intermediate coupling
($\lambda \approx 1$ to $\lambda \approx 2.5$ in this case)
between the two regions above. As $\lambda$ increases in this
region, a decrease in the PE and an increase in KE indicates the
localization of the polaron.
\end{enumerate}

\subsubsection{Variation with Phonon Frequency}
Now let us consider the way in which the properties of the
Holstein polaron are affected by altering the value of the
dimensionless phonon frequency $\bar{\omega}$, as defined by
Eq.~\ref{MHomega}. This quantity (also known as the {\it
adiabatic ratio}) is often used as a parameter in analytical
approaches to the polaron problem: the {\it adiabatic regime} is
defined as the case when $\bar{\omega}<1$, and the {anti-adiabatic
regime} as $\bar{\omega} >1$. With this in mind, we present below
the properties of the one-dimensional Holstein polaron for
$\bar{\omega}=0.5$, $\bar{\omega}=1$, and $\bar{\omega}=3$.

The ground state energies for all three values of $\bar{\omega}$ are
presented together in FIG.~\ref{Eall} against $\lambda$. The PE
tends to the same ($\bar{\omega}$-independent) SCP result of
$E_{0}(0)=2\lambda$ as $\lambda\rightarrow\infty$. Notice that the
KE for intermediate and large values of $\lambda$ (mainly due to
that of the lattice vibrations) decreases as $\bar{\omega}$
increases. This affects the position of the start and end of the
transition region.  More precisely, as $\bar{\omega}$ increases:
\begin{enumerate}
\item The start of the smooth transition region moves to higher
$\lambda$. \item The transition region becomes broader (the start
of the small-polaron region is also shifted to larger $\lambda$).
\end{enumerate}
The values of $\lambda$ that mark the estimated start and end of
the transition region are shown in TABLE \ref{HOLtran}.  The
definition of $\lambda$, in Eq.~\ref{Plam}, is independent
of $\bar{\omega}$, and so naturally characterizes the three
regions defined above.

The variation of $N_{\rm ph}$ with $\lambda$ is presented in
FIG.~\ref{Nph} for each value of $\bar{\omega}$. Over all
couplings, $N_{\rm ph}$ decreases as the value of $\bar{\omega}$
increases. Loosely speaking, it is simply ``harder'' to create
phonons of higher frequency. The smooth transition from large to
small polaron is again visible in the results for $N_{\rm ph}$,
with the edges of the transition region occurring at the same
$\lambda$ as in the above results for $E_{0}(0)$.

\begin{table}
    \centering
      \begin{tabular}{|c|c|c|}
    \hline
    $\bar{\omega}$ & $\lambda$(end of large-polaron region)&
     $\lambda$(start of small-polaron region)\\
    \hline
    0.5 & $(1.1\pm 0.1)$ & $(2.0\pm 0.2)$\\
    \hline
    1 & $(1.3\pm 0.1)$ & $(2.7\pm 0.3)$ \\
    \hline
    3 & $(2.1\pm 0.1)$ & $(5.5\pm 0.6)$  \\
    \hline
\end{tabular}
\caption{The boundaries of the transition region for the
    one-dimensional Holstein model at different values of the
    (dimensionless) phonon frequency $\bar{\omega}$.
    The estimates are based on the QMC results for the energies
    and $N_{\rm ph}$ (observing the changing "trend" in plots of
    the deviation from the corresponding
    SCP and WCP result).}
     \label{HOLtran}
\end{table}

The QMC results for the inverse effective mass $m_{0}/m^{*}$, and
the isotope exponent on the effective mass $\alpha_{m^{*}}$, are
presented against $\lambda$ in FIGS.~\ref{minv}--\ref{iso},
for each value of $\bar{\omega}$.  We see from FIG.~\ref{minv}
that increasing $\bar{\omega}$ reduces the effective mass over the
entire range of $\lambda$. Both $m_{0}/m^{*}$ and $\alpha_{m^{*}}$
tend to the WCP solution as $\lambda$ becomes small, and to the
SCP result as $\lambda$ becomes large.  However, the results for
$\ln(m^{*}/m_{0})$ tend to the SCP solution at a slower rate than
for the other observables.

From all the results for the Holstein model above, we observe that
as the value of $\bar{\omega}$ increases, the curve for each
observable (PE, $N_{\rm ph}$, $m^{*}$, and $\alpha_{m^{*}}$)
against $\lambda$ moves closer to the line representing the SCP
result over the {\it entire} range of $\lambda$.  In other words,
the SCP prediction becomes more applicable as $\bar{\omega}$
increases.

\subsection{Screened Fr\"{o}hlich Interaction}
Here we consider the case of the one-dimensional screened
Fr\"{o}hlich model, Eq.~\ref{FMfsc}, at a fixed
dimensionless phonon frequency of $\bar{\omega}=1$.  We
investigate the way the polaron properties depend on the range
of the electron-phonon interaction by comparing the QMC results at
four different values of the screening length (shown in FIG.~\ref{forces}): $R_{\rm sc}\to 0$ (the short-range Holstein
interaction), $R_{\rm sc}=1$, $R_{\rm sc}=3$, and $R_{\rm
sc}\to\infty$ (the non-screened Fr\"{o}hlich interaction).
\begin{figure}
\includegraphics[scale=0.5, angle=-90]{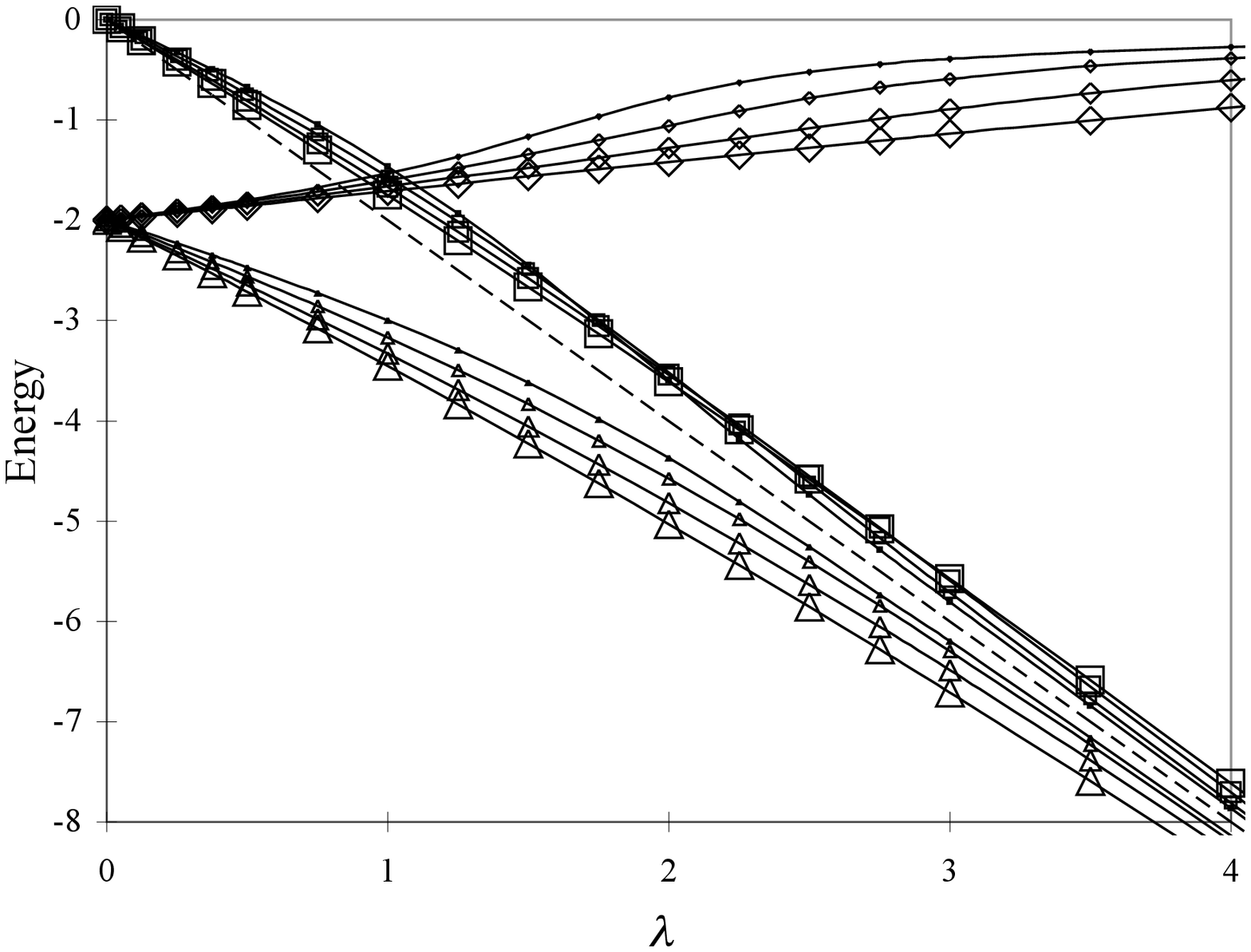}
\caption{The ground state energy $E_{0}(0)$ (triangles), potential energy
(squares), and kinetic energy (diamonds) of the one-dimensional screened
Fr\"{o}hlich model at $\bar{\omega}=1$, versus $\lambda$, for
screening lengths $R_{\rm sc}=0,1,3,\infty$ (increasing size of symbols). The curves for $E_{0}(0)$
(and PE) tends to the same strong coupling perturbation (SCP)
result of $E_{0}/t=-2\lambda$ (dashed line) as
$\lambda\to\infty$.  Note the crossing of the potential energy curves near $\lambda=2$.} \label{EF}
\end{figure}

\begin{figure}
\includegraphics[scale=0.5, viewport=0 300 800 800,clip]{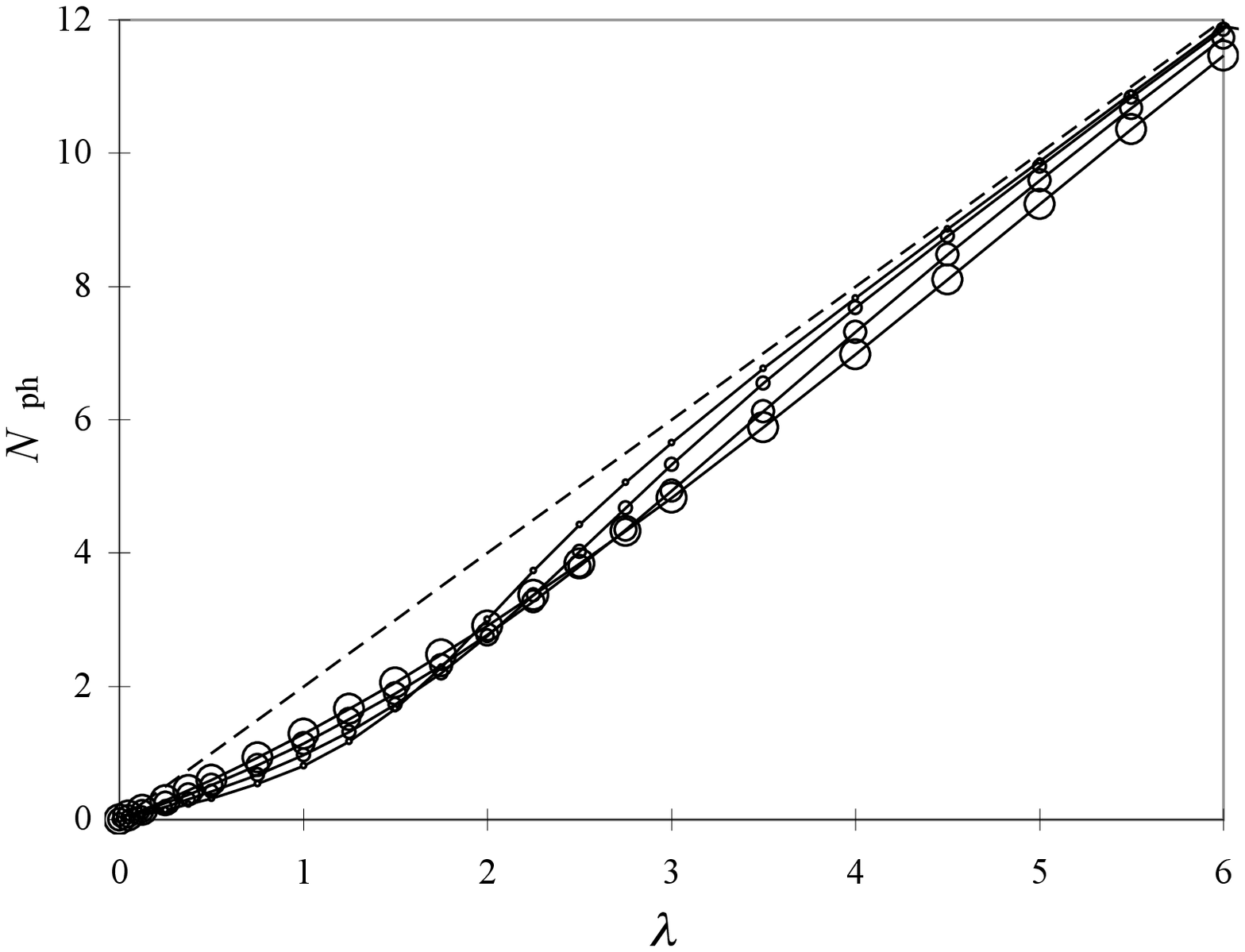}
\caption{The number of phonons in the polaron cloud $N_{\rm ph}$
versus $\lambda$ for screening lengths $R_{\rm sc}=0,1,3,\infty$ (increasing size of circles)  at $\bar{\omega}=1$.  The
curves tend to the same SCP result of $N_{\rm
ph}=z\lambda/\bar{\omega}$ (dashed line) as $\lambda\to\infty$.}
\label{NphF}
\end{figure}

\begin{figure}
\includegraphics[scale=0.5, angle=-90]{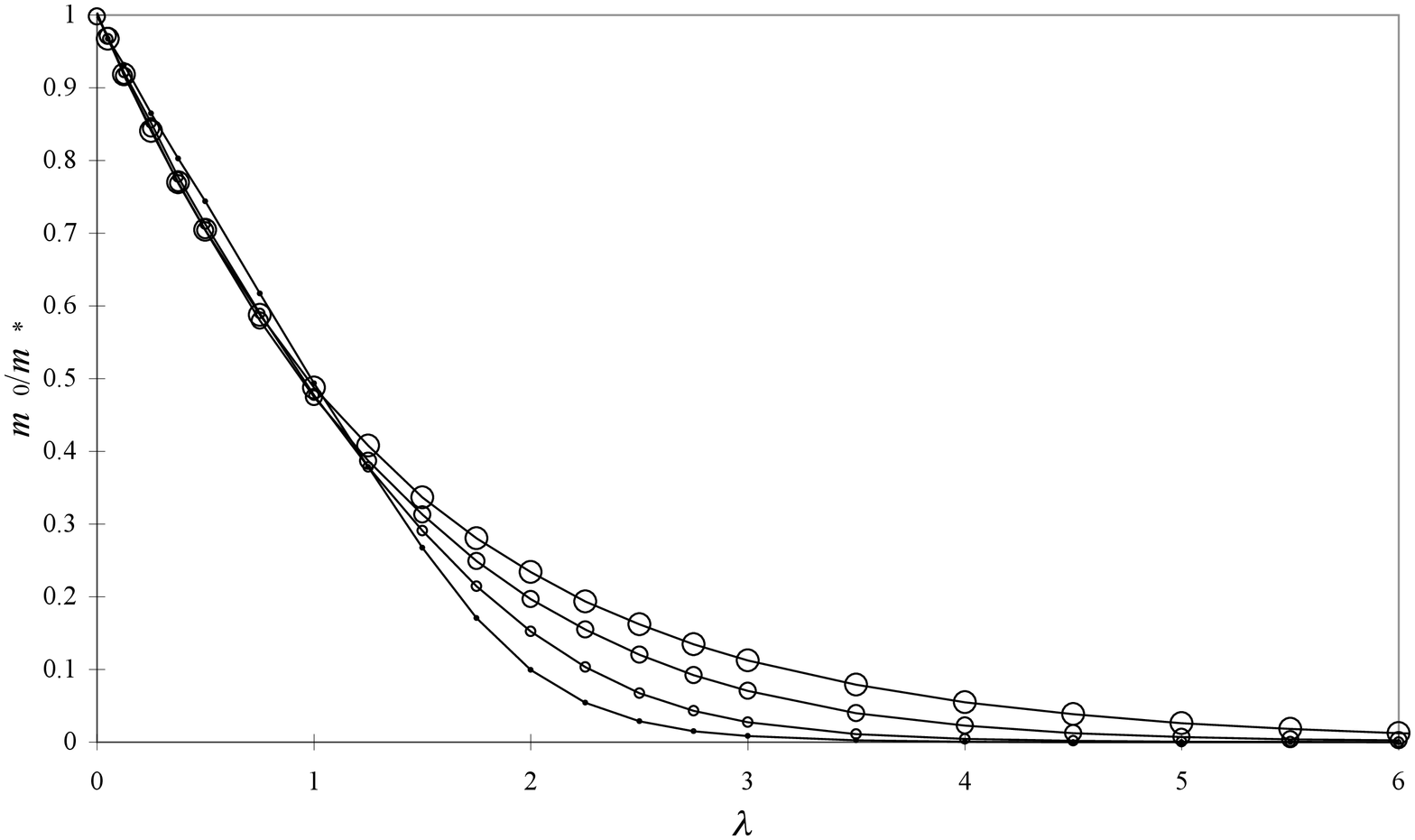}
\caption{The inverse effective mass $m_{0}/m^{*}$ for screening lengths $R_{\rm sc}=0,1,3,\infty$ (increasing size of circles) versus $\lambda$ at fixed $\bar{\omega}=1$.  For
weak coupling ($\lambda < 1$) the Holstein large-polaron has a
slightly smaller $m^{*}$ than the long-range interactions.}
\label{minvF}
\end{figure}

\begin{figure}
\includegraphics[scale=0.5, angle=-90]{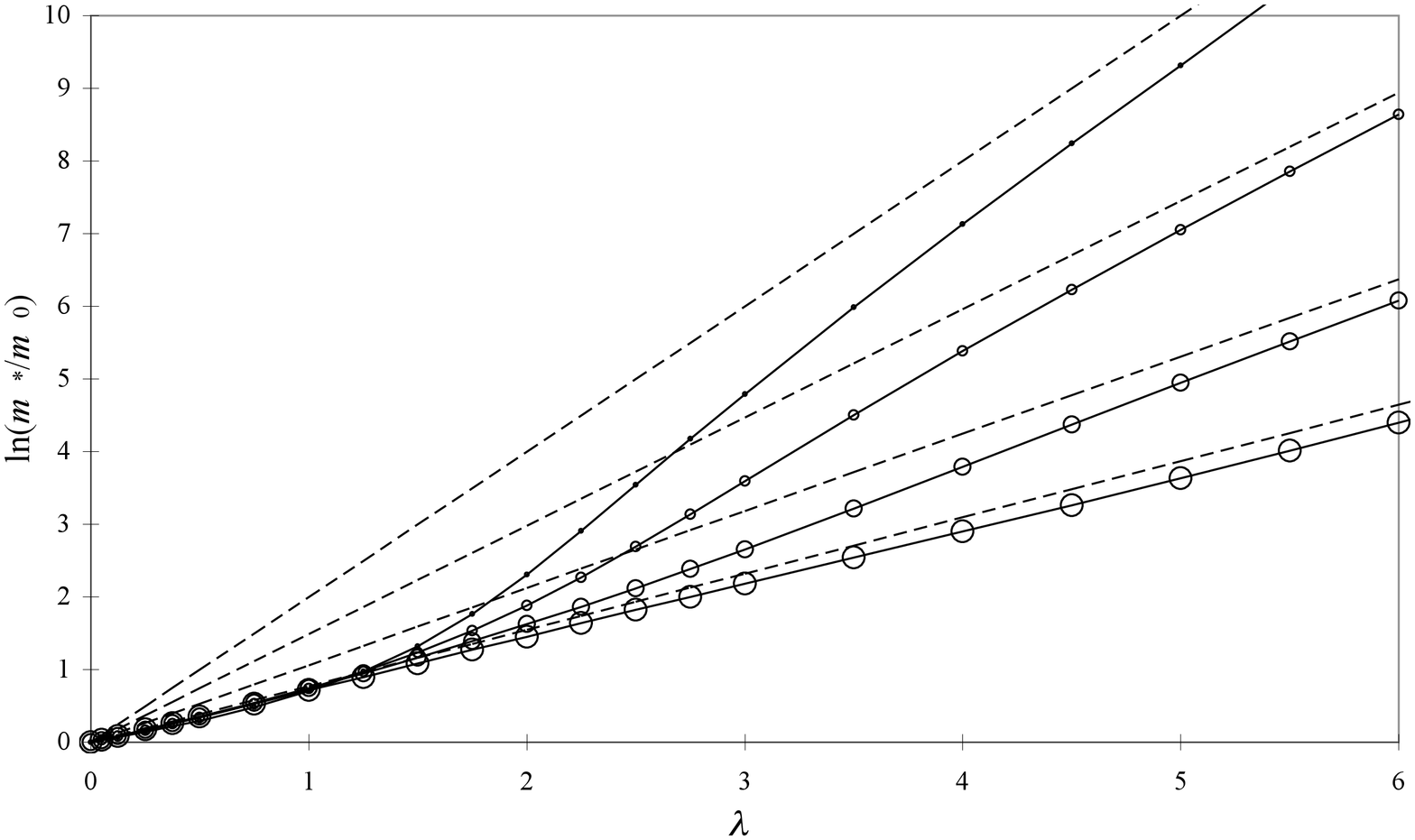}
\caption{The logarithm of the effective mass for screening lengths $R_{\rm sc}=0,1,3,\infty$ (increasing size of circles) versus $\lambda$ at $\bar{\omega}=1$.
At intermediate and strong coupling, decreasing the value of
$R_{\rm sc}$ dramatically increases the effective mass. The curves
tend to the SCP result (dashed lines) at a slower rate than
$E_{0}(0)$ and $N_{\rm ph}$.} \label{mF}
\end{figure}

\begin{figure}
\includegraphics[scale=0.5, angle=-90]{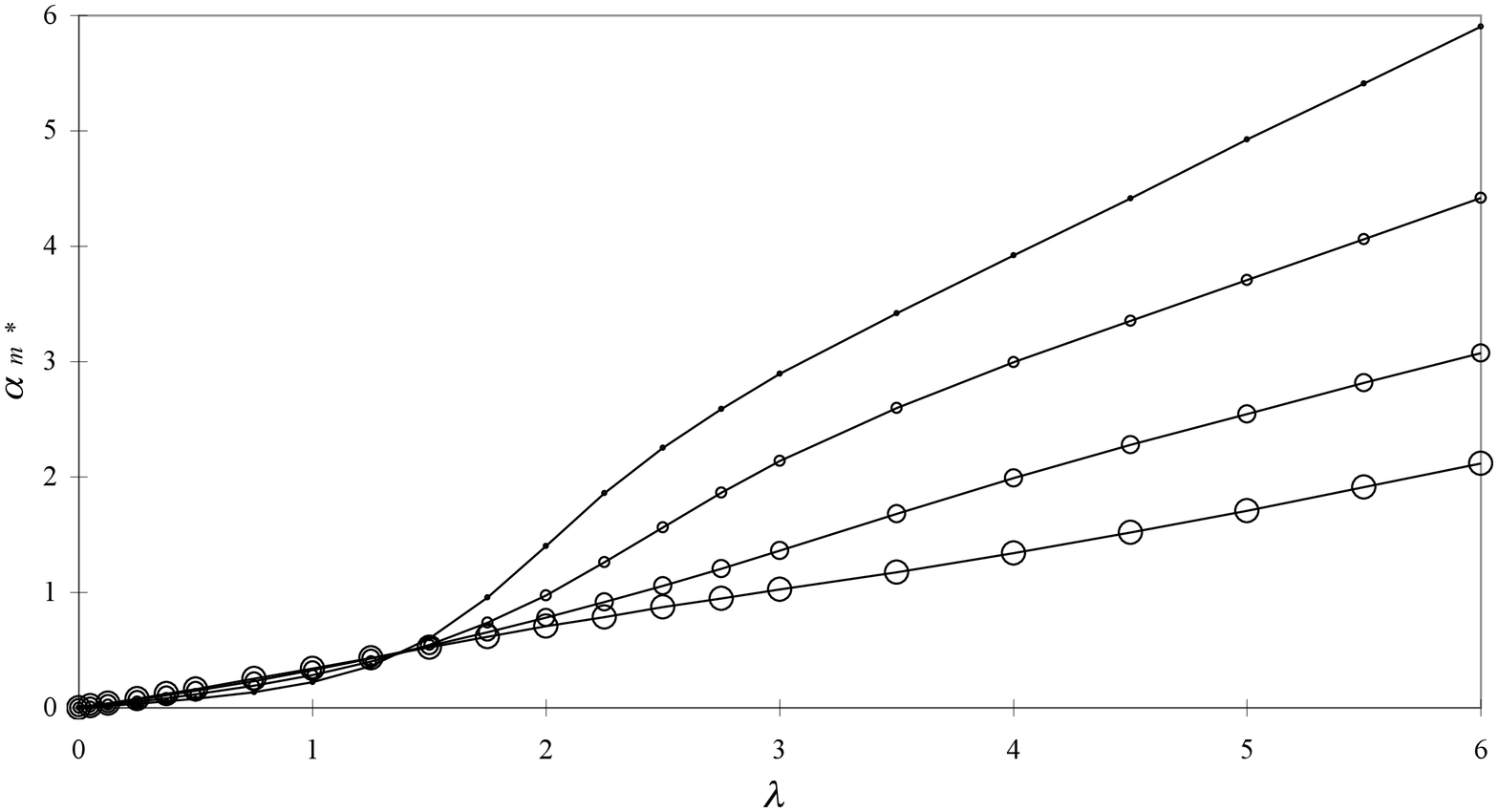}
\caption{The isotope exponent on the effective mass
$\alpha_{m^{*}}$ for screening lengths $R_{\rm sc}=0,1,3,\infty$ (increasing size of circles) versus $\lambda$  at
$\bar{\omega}=1$.} \label{isoF}
\end{figure}

The QMC results for the energy ($E_{0}(0)$, PE, and KE) and the
number of phonons in the polaron cloud $N_{\rm ph}$ are presented
against $\lambda$ in FIGS.~\ref{EF} and \ref{NphF}
respectively. They are in excellent agreement with the WCP results
at small $\lambda$ (not shown), and tend to the {\it same} (that
is, $\gamma$-independent) SCP result as
$\lambda\rightarrow\infty$. One can see that, as the value of
$R_{\rm sc}$ increases, the KE increases less rapidly with
$\lambda$, and so:
\begin{enumerate}
    \item  The start of the transition region shifts to higher $\lambda$.
    \item  The transition region becomes broader (in $\lambda$).
\end{enumerate}
This behavior is similar to that found for the Holstein model with
increasing $\bar{\omega}$. In the present case, with the phonon
frequency $\bar{\omega}$ fixed, the above effect is due only to
changing the shape of the electron-phonon interaction force
$f_{\bf m}({\bf n})$. TABLE \ref{LRtab} shows the values of
$\lambda$ that mark the estimated start and end of the transition
region for each value of $R_{\rm sc}$. From the $N_{\rm ph}$
results in FIG.~\ref{NphF}, we notice that increasing the
value of $R_{\rm sc}$ has the effect of {\it increasing} $N_{\rm
ph}$ at small $\lambda$, and the opposite effect of {\it
decreasing} $N_{\rm ph}$ at large $\lambda$.  That is, the order
that $N_{\rm ph}$ appears (with increasing $R_{\rm sc}$) in the
large polaron region is the reverse of the order in the small
polaron region.

\begin{table}[tbp]
    \centering
\begin{tabular}{|c|c|c|}
    \hline
    Interaction Model &  $\lambda$(start of transition)&
     $\lambda$(end of transition) \\
    \hline
    Holstein ($R_{\rm sc}\to 0$)&  $(1.1\pm 0.1)$ & $(2.0\pm 0.2)$\\
    \hline
    Screened Fr\"{o}hlich ($R_{\rm sc}=1$)  & $(1.7\pm 0.1)$ & $(3.8\pm
    0.4)$ \\
    \hline
    Screened Fr\"{o}hlich ($R_{\rm sc}=3$)  & $(2.2\pm 0.1)$ & $(4.7\pm
    0.4)$ \\
    \hline
    Non-screened Fr\"{o}hlich ($R_{\rm sc}\to\infty$)  & $(2.9\pm
    0.2)$ & $(6.3\pm 0.5)$ \\
    \hline
\end{tabular}
         \caption{The boundaries of the transition region for the
         one-dimensional screened Fr\"{o}hlich interaction,
         at various screening lengths $R_{\rm sc}$ (measured in units
         of the lattice constant). The estimates are
         based on the QMC results for the energy and $N_{\rm
         ph}$.}
    \label{LRtab}
\end{table}

The results for the effective mass are presented in FIGS.~\ref{minvF} and \ref{mF}, and the isotope exponent
results in FIG.~\ref{isoF}, against $\lambda$ for the same
four values of $R_{\rm sc}$. We can see that for each value of
$R_{\rm sc}$ the QMC results tend to the ``model dependent'' SCP
result (of $\ln(m^{*}/m_{0})=2\gamma\lambda/\bar{\omega}$ and
$\alpha_{m^{*}}=\gamma\lambda / \bar{\omega}$, where the value of
the mass enhancement factor $\gamma$ is given in TABLE
\ref{table1} for each $R_{\rm sc}$). This model-dependency is in
contrast to the above results for $E_{0}(0)$ and $N_{\rm ph}$,
which are $\gamma$-independent in the limit
$\lambda\rightarrow\infty$.

An important observation is evident from the plot of
$\ln(m^{*}/m_{0})$ against $\lambda$ shown in FIG.~
\ref{mF}. At intermediate and large couplings (that is,
in the transition and small polaron regions), altering the value
of $R_{\rm sc}$ has a {\it dramatic} effect on the effective mass.
For example, the non-screened Fr\"{o}hlich polaron is over
$10^{3}$ times ``lighter'' than the Holstein polaron at
$\lambda=4$, and over $10^{4}$ times lighter at $\lambda=5$.

The isotope exponent $\alpha_{m^{*}}$ in FIG.~\ref{isoF}
shows a strong dependence on the range of electron-phonon
interaction (as well as on $\bar\omega$ and $\lambda$) in the
(physically most realistic) intermediate values of coupling.  This
is important, as experimental measurement (by Zhao {\it et al.})
of the exponent of the isotope exponent on the effective
supercarrier mass along the ${\rm CuO_{2}}$ planes, $m_{ab}^{*}$,
in the material ${{\rm La}_{2-x}{\rm Sr}_{x}{\rm CuO}_{4}}$ shows
a large value of $\alpha_{m^{*}}^{(ab)}=1.9(2)$ in the deeply
underdoped regime ($x=0.06$), and a much smaller value of
$\alpha_{m^{*}}^{(ab)}=0.46(5)$ for optimal doping ($x=0.15$)
\cite{zhao,Zhao2}.  Both the magnitude and radius of the electron-phonon
interaction should decrease with doping due to screening.  These experimental results show that the former effect is more significant.

Now let us consider the large-polaron region. The QMC results for
$m^{*}$ and $\alpha_{m^{*}}$ tend to the WCP results for all
$R_{\rm sc}$ (not shown) as $\lambda\rightarrow 0$.  As was the
case for $N_{\rm ph}$, we see from FIGS.~\ref{minvF} and
\ref{isoF} that the order that $m^{*}$ and $\alpha_{m^{*}}$
appear in the large polaron region (with increasing $R_{\rm sc}$)
is the opposite to that in the small polaron region.  As can be
seen in FIG.~\ref{minvF}, the decrease of $m_{0}/m^{*}$
with $\lambda$ is approximately linear for the Holstein model, and
approximately exponential for the other screening lengths.  In
fact, the effective mass of the  Fr\"{o}hlich large-polaron is
{\it larger} (up to approximately $10\%$) than the effective mass
of the (short-range) Holstein large-polaron.

It is apparent from the above results for the screened
Fr\"{o}hlich model that as $R_{\rm sc}$ increases (from Holstein
to Fr\"{o}hlich), the QMC results  move, in general, closer to the
SCP prediction over the {\it entire} range of $\lambda$. That is,
the SCP prediction becomes generally more applicable as the range
of interaction increases (as well as with increasing
$\bar\omega$).
This fact allows us to define the intermediate region of the coupling
strength and of the adiabatic ratio as a ``small polaron'' regime for any
realistic-range e-ph interaction, that is the regime which is well
described by the small polaron theory based on the Lang-Firsov transformed
Hamiltonian averaged over phonons.

\subsection{Comparison with other approaches}

\begin{figure}
\includegraphics[scale=0.5, angle=-90]{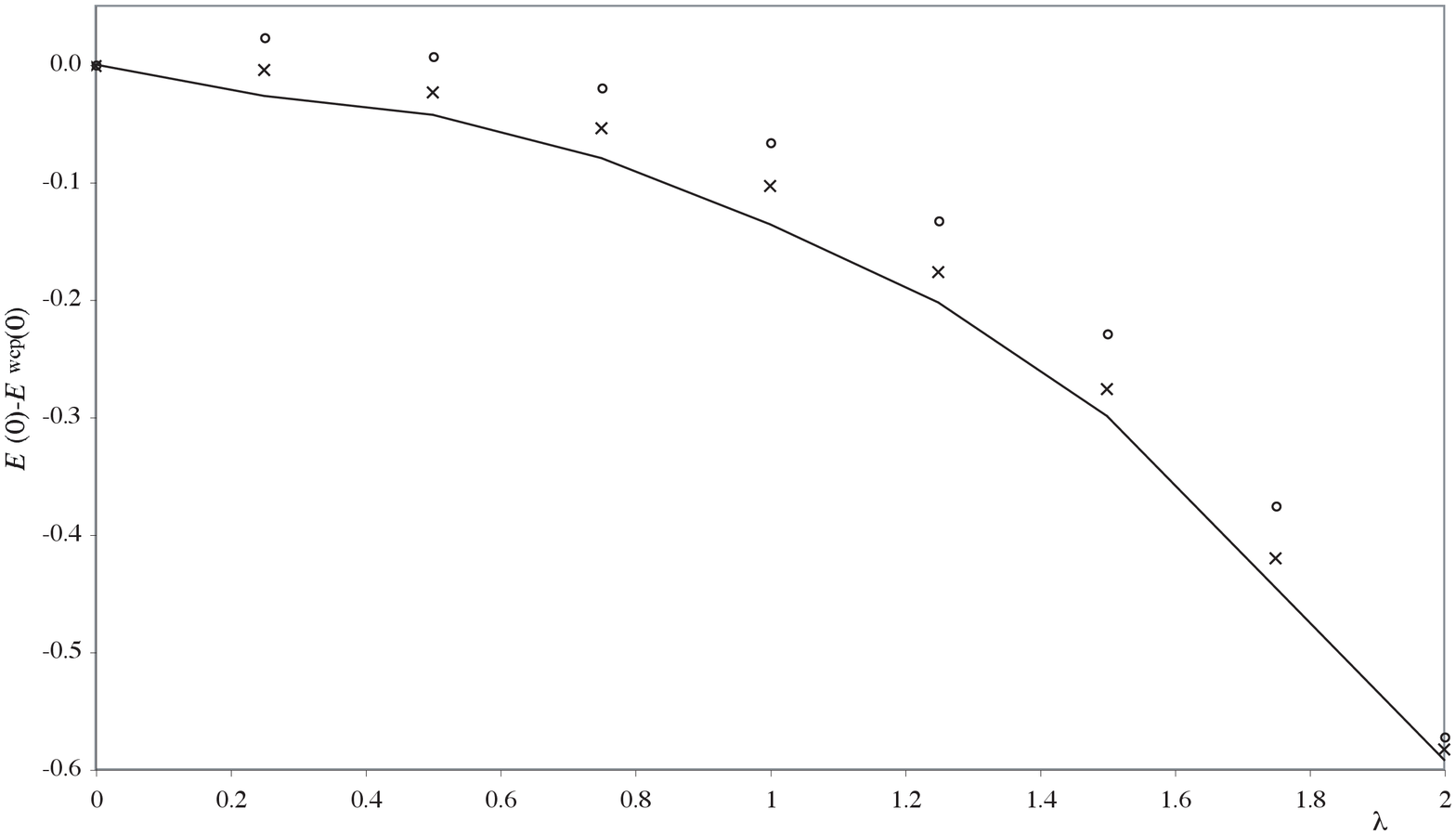}
    \caption{1D Holstein polaron ground-state energy for $\bar{\omega}= 1.0, R=0$ (crosses) compared with bosonic QMC (circles, Ref.~\onlinecite{Hoh}) and exact diagonalization energies (solid line, Ref.~\onlinecite{Marsiglio})  at
    $\bar{\omega}= 1.0$.  The energy in  weak-coupling perturbation theory (\ref{WCNe2}) has been subtracted.}
    \label{Ecomp}
\end{figure}

\begin{figure}
\includegraphics[scale=0.5, angle=-90]{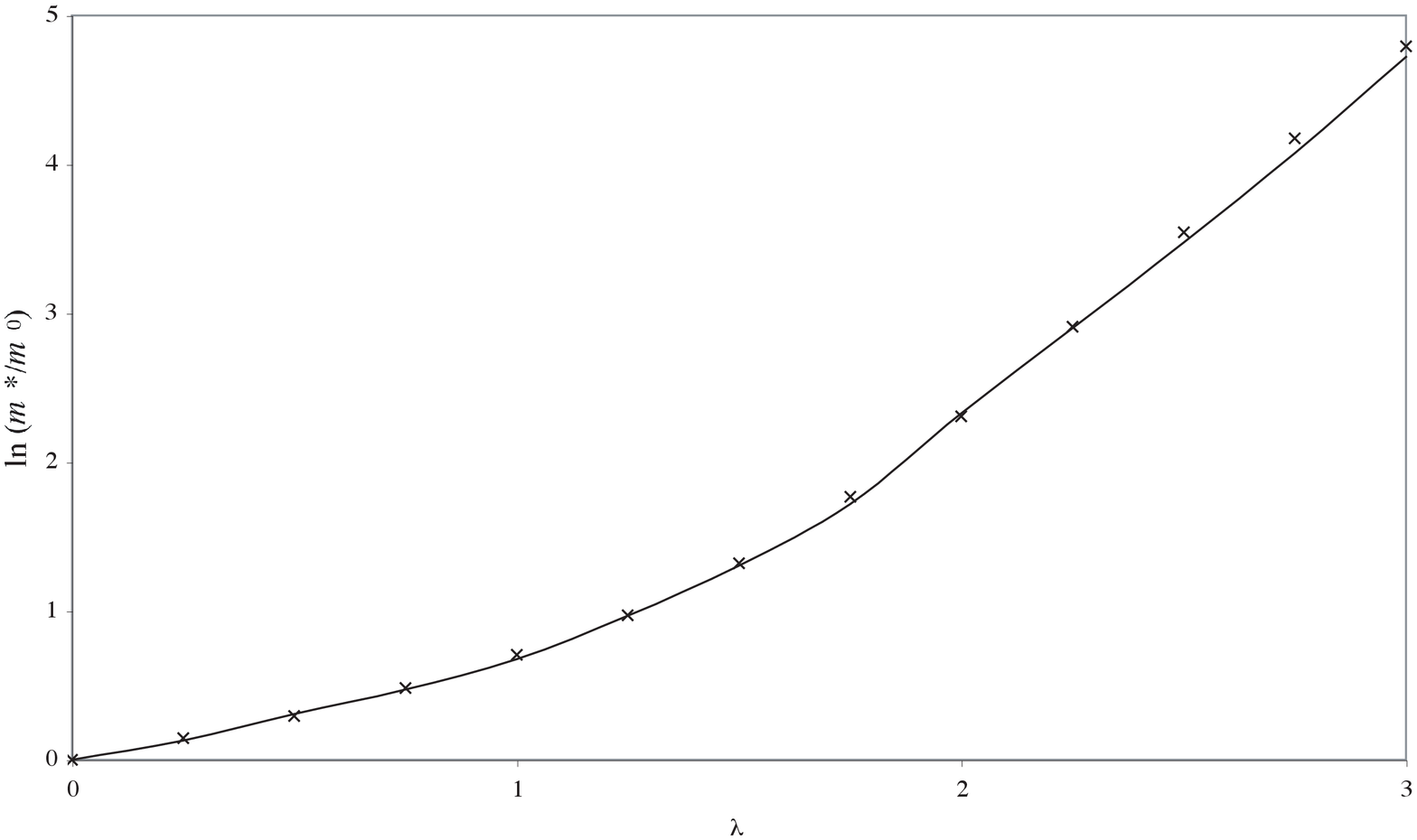}
    \caption{1D Holstein polaron effective mass for $\bar{\omega}= 1.0, R=0$ (crosses) compared with  variational results  (solid line, Ref.~\onlinecite{trugman}).}
    \label{mcomp}
\end{figure}

As a test of our method, we compare our results with those of other authors.
Our main and original computations are for finite values of $R$; however, the majority of published work on lattice polarons relates to the Holstein ($R=0$) interaction. FIG.~\ref{Ecomp} compares our ground state energy for $\bar{\omega}= 1.0, R=0$ (compare FIG.~\ref{E}) with that obtained by other authors.  To highlight the differences, the weak coupling approximation (\ref{WCNe2}) is subtracted from the energy.  The figure compares our results with the QMC data of Hohenadler \emph{et al.}\cite{Hoh} (based on a bosonic path integral, evaluated at inverse temperature $\bar\beta=10$) and exact diagonalization results for small clusters\cite{Marsiglio}.  Different computations yield similar values of the ground state energy, but our QMC energies are closer to the the exact 
diagonalization results; our calculations have been carried out at a lower temperature, $\bar\beta=25$.  FIG.~\ref{mcomp} shows good agreement between our results for the effective mass (compare FIG.~\ref{m}) and variational calculations of Bon\v{c}a  \emph{et al.}\cite{trugman}  (Note that the $\lambda$ in that work corresponds to $\sqrt{2\omega t\lambda}$ in our notation.)
We also reinforce earlier conclusions on the dependence of effective mass with interaction range\cite{Korn2, feh3} by interpolation between the Holstein and Fr\"{o}hlich limits.
\section{Conclusions}                   \label{sect7}               
The general aim of this work was to investigate the way in which
the {\it range} of electron-phonon interaction governs the
physical properties of the (single) lattice polaron.  The
understanding of this is of considerable current interest because
of the increasing amount of experimental evidence suggesting that
polarons are present in high-temperature superconducting and
colossal magnetoresistance materials.

Perturbation approaches are used in the limits of strong and weak
electron-phonon coupling strength $\lambda$.  However, in general
these do not provide an acceptable description in the (physically
most realistic) intermediate coupling range $\lambda\approx 1$. We
have performed an extensive Monte Carlo study of the ground state
properties for the screened Fr\"{o}hlich polaron, equation
(\ref{FMfsc}), in one dimension, over a wide range of coupling.

We have used path-integral quantum Monte Carlo, in which the phonon
degrees of freedom are analytically integrated out, leaving only
the electron coordinates to be simulated. The use of a path
integral with {\it twisted} boundary conditions allowed us to extract dynamic properties {\it
directly} from the simulations. There
were no (systematic) errors due to finite size or finite time-step.

The properties measured were the ground state energy $E_{0}(0)$,
the number of phonons in the polaron cloud $N_{\rm ph}$, the
effective mass $m^{*}$, and the isotope exponent on the effective
mass $\alpha_{m^{*}}$. The QMC results were always found to tend
to the weak-coupling perturbation (WCP) predictions for
$\lambda\rightarrow 0$, and to the strong-coupling perturbation
(SCP) predictions for $\lambda\rightarrow\infty$.

The screened Fr\"{o}hlich polaron was studied for various values
of the screening length $R_{\rm sc}$ (which essentially controls
the range of the electron-phonon interaction): $R_{\rm
sc}\rightarrow 0$ (on-site, Holstein interaction), $R_{\rm sc}=1$,
$R_{\rm sc}=3$, and $R_{\rm sc}\rightarrow\infty$ (long-range,
non-screened Fr\"{o}hlich interaction). For each value of $R_{\rm
sc}$, we determined the variation of the above observables with
$\lambda$, at a fixed phonon frequency of $\bar{\omega}=1$. The
main findings are summarized below:
\begin{enumerate}

\item We observe the presence of a self-trapping transition for
{\it all} values of $R_{\rm sc}$.  In each case, the following
three regions are identified:
\begin{enumerate}
    \item  The large-polaron region at weak coupling, in which the
    behavior of the system is accurately described by WCP theory.
    This region is characterized by delocalized, band-electron-like states.

    \item  The small-polaron region at strong coupling,
     in which the behavior is accurately
    described by SCP theory.  This region is characterized by
    localized (``self-trapped'') polaronic states.

    \item  The transition region between the two, at intermediate coupling.
       We observe a smooth crossover from large to small polaron
       in {\it all} the observables measured.
\end{enumerate}

\item The transition region boundaries depend on the range of
interaction.  As the value of $R_{\rm sc}$ increases
\begin{enumerate}
    \item  The start of the transition region (the point at which it
    becomes energetically favorable for localized states to exist)
    shifts to higher $\lambda$.
    \item  The transition region becomes broader (the start of the
    small polaron region also shifts to higher $\lambda$).
    The small-polaron region starts when the kinetic energy is much
    smaller than the potential energy.
    \item  The values of the observables
    (PE, $N_{\rm ph}$, $m^{*}$, and $\alpha_{m^{*}}$) generally move
    closer to the corresponding SCP result over the {\it entire}
    range of $\lambda$.
\end{enumerate}

\item In the large polaron region, the effective mass for
long-range electron-phonon interaction ($R_{\rm sc}>1$) is found
to be up to approximately $10\%$ larger than that for the Holstein
interaction ($R_{\rm sc}\to 0$).

\item We observe large variations in the isotope exponent on the
effective mass $\alpha_{m^{*}}$ in the (physically most realistic)
intermediate coupling regime (with changing $R_{\rm sc}$ and
$\lambda$, as well as $\bar\omega$).  This is encouraging, as
experimental observation shows large variations in the isotope
exponent with the level of doping in high-$T_{\rm c}$ materials.

\item Reducing the range of electron-phonon interaction
dramatically increases the effective mass for intermediate and
large values of coupling (that is, in the transition and
small-polaron regions). In comparison, $E_{0}(0)$ and $N_{\rm ph}$
are only slightly affected by altering $R_{\rm sc}$.

\item We also study the dependence on phonon frequency
$\bar{\omega}$ of the Holstein polaron
($R_{\rm sc}\rightarrow 0$). As $\bar{\omega}$ increases
\begin{enumerate}
\item The transition region shifts to higher $\lambda$. 
\item The
transition region becomes broader. 
\item The observables (PE,
$N_{\rm ph}$, $m^{*}$, and $\alpha_{m^{*}}$) move towards the
corresponding SCP result over the entire range of $\lambda$.
\end{enumerate}

\item Increasing interaction range has a qualitatively similar effect to
increasing phonon frequency; compare for example FIGS.~\ref{m} and \ref{mF}
or FIGS.~\ref{iso} and \ref{isoF}.
\end{enumerate}

One can also  calculate the isotope effect on the whole polaron
band dispersion  applying the continuous-time quantum
Monte-Carlo algorithm \cite{alekorfut}. To deal with the electron
spectral function and high-energy excitations, involving phonon
shake-off,  measured in ARPES \cite{lan0,shencon,lan}, the QMC
algorithm has to include the off-diagonal paths, which remains a
challenging but solvable problem of our QMC simulations in the site representation. Also other methods as the momentum based QMC \cite{pro}, the numerical diagonalization of vibrating
clusters or the renormalisation group are able to calculate the spectral function. The method used here 
relies on a phonon gap, the errors 
being exponentially small in $\beta\hbar\omega$,  and is 
therefore more accurate for intermediate and large values of $\omega$. 
While
lower $\omega$ can be also simulated without major difficulty but with increased inverse temperature and therefore increased CPU time,  the parameters used cover the most physically relevant range\cite{alemot}.

\section*{Acknowledgments}
PES acknowledges support from an EPSRC studentship, and ASA
acknowledges support of the Leverhulme Trust (London).

\bibliographystyle{apsrmp}

\end{document}